# Programming nonreciprocity and reversibility in multistable mechanical metamaterials

Gabriele Librandi[1], Eleonora Tubaldi[2,*], Katia Bertoldi[1,†]

[1] John A. Paulson School of Engineering and Applied Sciences,
Harvard University, Cambridge, MA 02138, USA.
[2] Mechanical Engineering Department,
University of Maryland, College Park, MD 20742, USA.
* etubaldi@umd.edu
† bertoldi@seas.harvard.edu

**Abstract**

Nonreciprocity can be passively achieved by harnessing material nonlinearities. In particular, networks of nonlinear bistable elements with asymmetric energy landscapes have recently been shown to support unidirectional transition waves. However, in these systems energy can be transferred only when the elements switch from the higher to the lower energy well, allowing for a one-time signal transmission. Here, we show that in a mechanical metamaterial comprising a 1D array of bistable arches nonreciprocity and reversability can be independently programmed and are not mutually exclusive. By connecting shallow arches with symmetric energy wells and decreasing energy barriers, we design a reversible mechanical diode that can sustain multiple signal transmissions. Further, by alternating arches with symmetric and asymmetric energy landscapes we realize a nonreciprocal chain that enables propagation of different transition waves in opposite directions.

**Key words:** Nonreciprocity - transition waves - shallow arches

Nonreciprocity – asymmetric transmission of energy between any two points in space – is receiving increasing interest in many areas of physics (*1, 2*), including optics (*3, 4*), electromagnetism (*5, 6*), elasticity (*7, 8*) and acoustic (*9–12*). Focusing on elastic systems, nonreciprocity has been successfully exploited to realize selective signal transmission (*13–17*), logic elements (*18, 19*), direction-dependent insulators (*20, 21*), and switches (*22*). To achieve such remarkable behaviors both active and passive strategies have been proposed. On the one hand, nonreciprocity for linear waves has been obtained either by imparting a rotation to the medium (*23*) or by introducing activated materials with time-modulated properties in space and time (*24–26*) to break time reversal symmetry. On the other hand, nonreciprocity has also been demonstrated in passive media by harnessing nonlinear phenomena (*27,28*). In particular, mechanical metamaterials with two or more stable equilibrium states have recently emerged as a powerful platform to realize nonreciprocity, as they support only unidirectional transition wave propagation when comprising an array of bistable building blocks with asymmetric energy wells (*19, 29–31*). However, while this strategy is appealing for its simplicity and robustness, it typically leads to non-reversible wave propagation since the systems release a net amount of energy upon propagation of the pulses and need to be manually 'recharged' (*i.e.* all elements need to be reset to their higher energy well) to sustain a second wave.

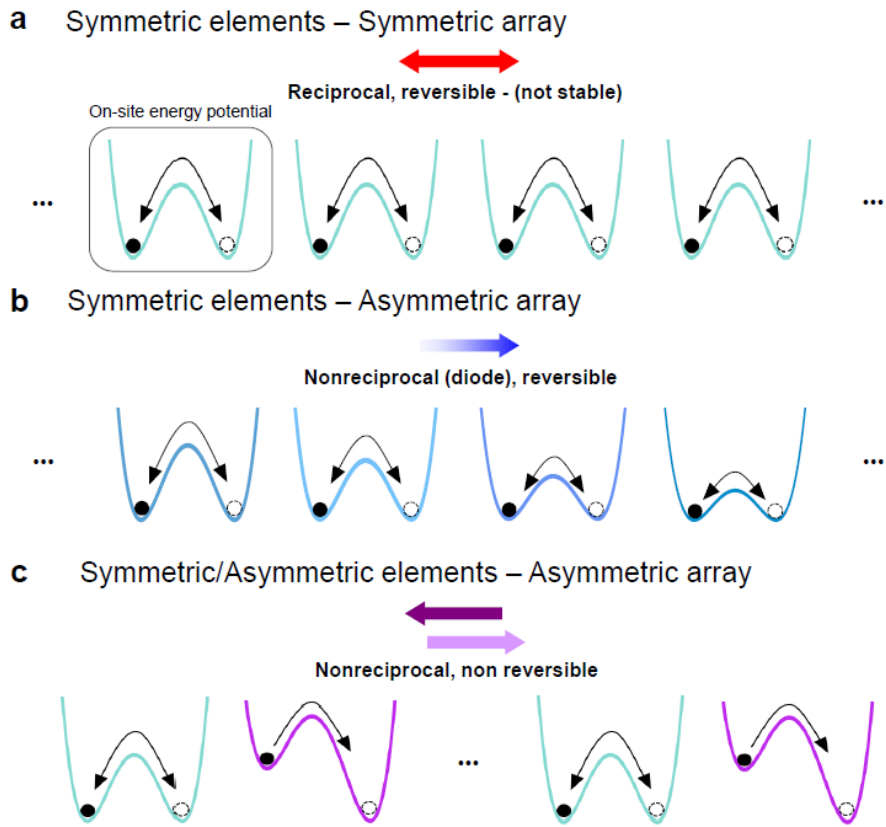

Figure 1: **Programming nonreciprocity and reversibility. a**, Signal propagation in a chain comprising identical bistable elements with symmetric energy wells is reciprocal and reversible, but not stable. **b**, Signal propagation in a chain comprising bistable elements with symmetric energy wells, but decreasing energy barriers is nonreciprocal and reversible. **c**, Signal propagation in a chain comprising bistable elements with both symmetric and asymmetric energy wells is nonreciprocal and non reversible.

Here, we demonstrate realization of a multistable mechanical metamaterial for which nonreciprocity and reversibility can be independently programmed. Such control of the dynamic response is made possible by the rich and highly tunable behavior of shallow arches, as their energy landscape can be easily adjusted to exhibit target energy barriers as well as symmetric or asymmetric wells. We first show that chains comprising identical arches with symmetric energy wells support the propagation of nonlinear pulses that sequentially switch the elements to their inverted stable configuration. However, while such signal propagation is reciprocal and reversible, it is not stable (Fig. 1a). Then, we demonstrate that by carefully designing the arches and their arrangement to break symmetry either at the structural or element level we can enable not only stable propagation of the signal, but also a wide range of nonreciprocal behaviors. For example, a reversible diode can be created by connecting shallow arches with symmetric but modulated on-site energy potentials (Fig. 1b). Further, a tunable 1D nonreciprocal chain, which enables propagation of different transition waves in opposite directions, can be obtained by alternating shallow arches with symmetric and asymmetric energy potentials (Fig. 1c). As such, our work open avenues for the design of the next generation of nonlinear structures and devices with robust, nonreciprocal elastic wave-steering capabilities.

We consider 1D chains comprising $N$ shallow arches connected via rotating hinges that impose continuity of rotations between adjacent elements. All arches have end-to-end distance $L$ = 120 mm and are made of spring steel shims with thickness $h$ = 0.3048 mm, width $b$ = 10 mm, length $l \in [103.1, 105.0]$ mm, volumetric density $\rho$ = 7850 kg/m$^3$ and Young's modulus $E$ = 170 GPa (see Supplementary Information for details). To excite the system, we move with an indenter the midpoint of either the first or last arch in the array at a constant speed $\alpha$ = 15 mm/s. We then monitor the response of the chain with a high-speed camera and track the position of the central point of the $j$-th arch, $w_j(L/2, t)$, as a function of time $t$ (see Supplementary Information for details).

We start by focusing on an array comprising three arches (*i.e.* $N$ = 3) with rise $e_j = w_j(L/2, t = 0)$ = 12.4 mm ($j$ = 1,2,3) realized by elastically buckling flat metallic shims of length $l$ = 105 mm (see Supplementary Information for details). The results reported in Figs. 2a and b for a test in which the indenter acts on the leftmost arch show two key features. First, as recently observed for individual hinged arches under displacement control (*32*), the indenter makes the leftmost arch snap to its symmetric stable configuration through the activation of the first asymmetric deformation mode (see Fig. 2a). Second, and most important, this reconfiguration does not remain localized as the energy released by the arch upon snapping is transmitted to the neighboring element through the rotation of the hinges. As a result, the snapping of the first arch triggers a cascade of snapping events that sequentially switches the other two elements to their symmetric stable configuration (see Movie S1). This response is fully reciprocal and reversible since actuating the first or the last arch (*i.e.* left-to-right *v.s.* right-to-left), from the top or the bottom (*i.e.* up-to-down *v.s.* down-to-up) always produces the same dynamic behavior (see Supplementary Information and Fig. S8 for details). However, if a fourth arch is added to the chain (*i.e.* for $N = 4$), the input provided by the indenter is not sufficient to generate a signal that switches all the elements of the lattice (Figs. 2c and d, see Movie S2).

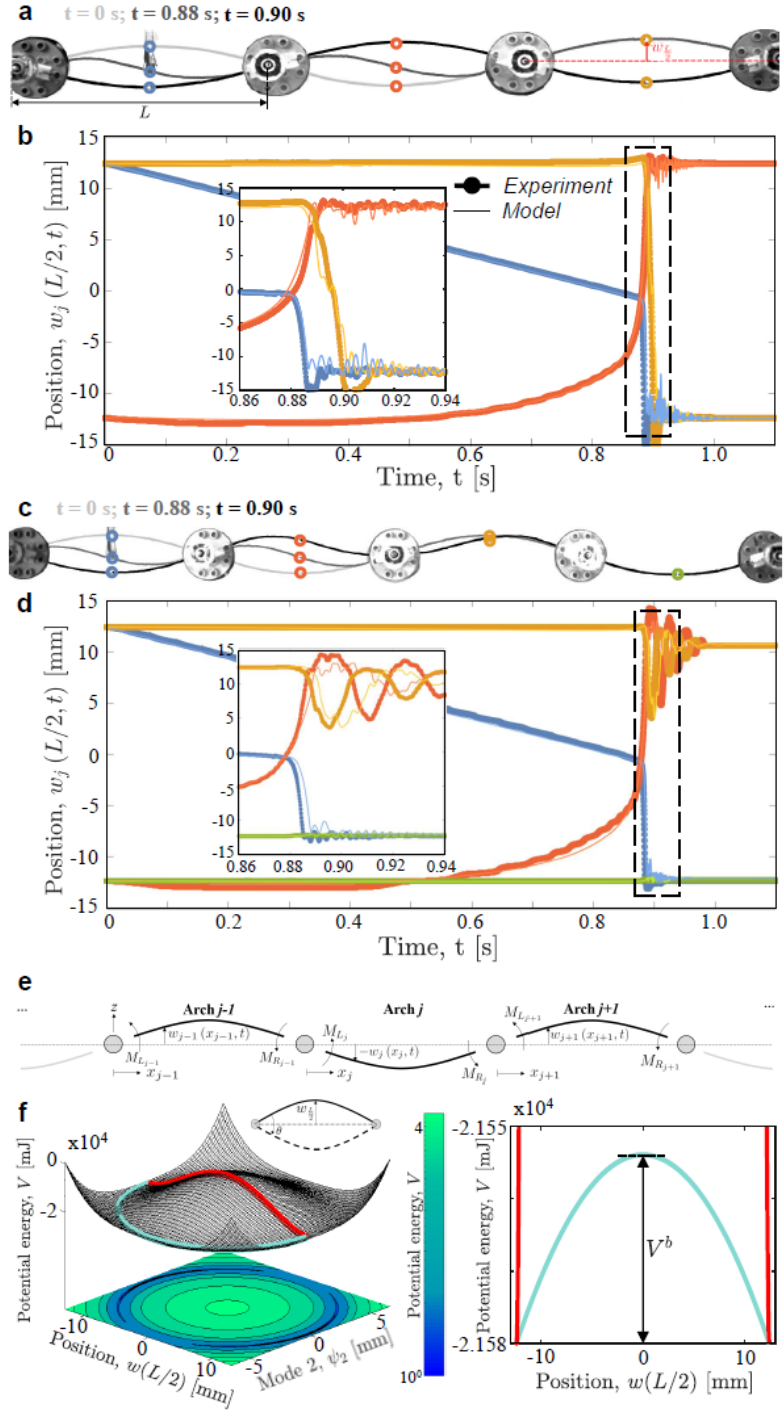

Figure 2: **Symmetric elements - Symmetric array. a-b**, Array comprising three identical arches with rise $e_j$ = 12.4 mm ($j$ = 1,2,3) and symmetric energy wells. **a** Snapshots at $t$ = 0s, 0.88s, 0.90s and **b**, evolution of the positions of the midpoints of the arches, $w_j(L/2, t)$, for a test in which the indenter acts on the leftmost arch. Thick-dotted and thin lines correspond to experimental and numerical results, respectively. **c-d**, Same as a-b but for an array comprising four arches. **e**, Schematic of the system. **f**, On-site energy potential for an elastically deformed shallow arch. The red line indicates a deformation path along which only the first symmetric mode is activated. The cyan line corresponds to the minimum energy path.

To get a deeper understanding on the snapping signal transmission through the chain, we establish a numerical model. We focus on the $j$-th arch, use Kirchhoff plate theory (33) to describe its response (32, 34–39) and impose continuity of rotations between neighboring elements (40, 41)

$$\left.\frac{\partial w_{j-1}}{\partial x_{j-1}}\right|_{x_{j-1}=L} = \left.\frac{\partial w_j}{\partial x_j}\right|_{x_j=0}, \tag{1}$$

where $x_j \in [0, L]$ represents the local axial coordinates and $w_j(x_j, t)$ denotes the time-dependent profile of the $j$-th arch. Importantly, the constraints described by Eq. (1) introduce concentrated moments at both ends of the $j$-th arch, $M_{Lj}$ and $M_{Rj}$ (Fig. 2e), and these satisfy

$$M_{R_{j-1}} = -M_{L_j}, \quad M_{R_j} = -M_{L_{j+1}}. \tag{2}$$

It follows that the response of a chain comprising arches can be described by

$$\rho A \frac{\partial^2 w_j}{\partial t^2} + \beta \frac{\partial w_j}{\partial t} + EI \left( \frac{\partial^4 w_j}{\partial x_j^4} - \frac{d^4 w_{0j}}{dx_j^4} \right) + p_j \frac{\partial^2 w_j}{\partial x_j^2} + Q_1 \delta_{1j} + M_{Lj}(1-\delta_{1j}) + M_{Rj}(1-\delta_{Nj}) = 0 \quad \text{for} \quad j = 1, ..., N \tag{3}$$

where $A$ and $I$ are the area and moment of inertia of the arches' cross section, $\rho$ and $E$ are the volumetric density and Young's modulus of the material, $\beta$ represents the viscous damping coefficient and $\delta_{1j}$ and $\delta_{Nj}$ are Kronecker delta functions. Moreover, $Q_1$ is the reaction force measured at the indenter and $w_{0j}$ and $p_j$ are the initial unstressed position of the midsurface and the midplane force produced by the stretching of the midsurface of the $j$-th arch, respectively (see Supplementary Information for more details). For our system $w_j(x_j, t)$ can be expressed as a series of sine functions (32)

$$w_j(x_j, t) = w_{0j}(x_j) + \sum_{n=1}^{N_t} \psi_{nj}(t) \sin\left(\frac{n\pi x_j}{L}\right). \tag{4}$$

Substitution of Eq. (4) into Eqs. (3) leads to a system of $N_t \times N$ coupled ordinary differential equations that we numerically solve to obtain the modal amplitudes $\psi_{nj}$. As shown in Figs. 2b and d, the numerical predictions are in very good agreement with the experimental results when choosing $N_t = 3$ and $\beta = 1.4 \cdot 10^{-6}$ kg/(m·s) (see Supplementary

Information for details and Movie S1) and capture both the propagation of the snapping signal through the entire chain for $N = 3$ and its arrest for $N = 4$.

To understand the absence of a stable propagation in the system comprising $N = 4$ elements, we focus on a single hinged arch and use Kirchhoff plate theory to determine its energy landscape when one of its ends is forced to rotate (see Supplementary Information for details). As observed in our experiments, we find that it is energetically more favorable for the arches to activate the first antisymmetric mode when snapping between the two stable states (see cyan path in Fig. 2f). However, despite the asymmetric deformation path, the on-site energy potential of the arches is symmetric and characterized by two wells of equal height at $w_{L/2} = \pm e_j$ separated by an energy barrier $V^b = 26$ mJ. As such, there is no net-release of energy when the arches snap between their two stable configurations and the stable propagation of the snapping signal is only possible in unrealistic systems without any form of dissipation (see numerical results for $\beta = 0$ in Fig. S9). To achieve stable wave propagation as well as to independently control reciprocity and reversibility, we then introduce asymmetry into the system both at the structural and arch level. To begin with, we build a 1D non-symmetric array by assembling elastically deformed shallow arches with monotonically decreasing rises (Fig. 3a and Table S1 for details). Since the energy barrier $V^b$ monotonically decreases as $e_j$ becomes smaller (Figs. 3b and c), the effect of dissipation can be counteracted by tuning the rises to make $\Delta V_j^b = V_{j-1}^b - V_j^b$ larger than the energy dissipated by the $j$-th arch during snapping. As shown in Fig. 3d, when the indenter excites the leftmost arch with the highest rise, a stable snapping wave propagates from left-to-right. Importantly, such wave propagation is reversible and non-reciprocal. Since all the arches have a symmetric energy landscape, the wave can be excited by snapping the first arch both up-to-down and down-to-up and there is no need to 'manually' recharge the system to propagate a new signal (*i.e.* the behavior is fully reversible - see Fig. S10). However, when the indenter excites the arch with the lowest rise the energy released upon its snapping is not enough to make the next arch to jump. As such, there is no wave propagation from right-to-left and the system acts as a mechanical diode (see Movie S3).

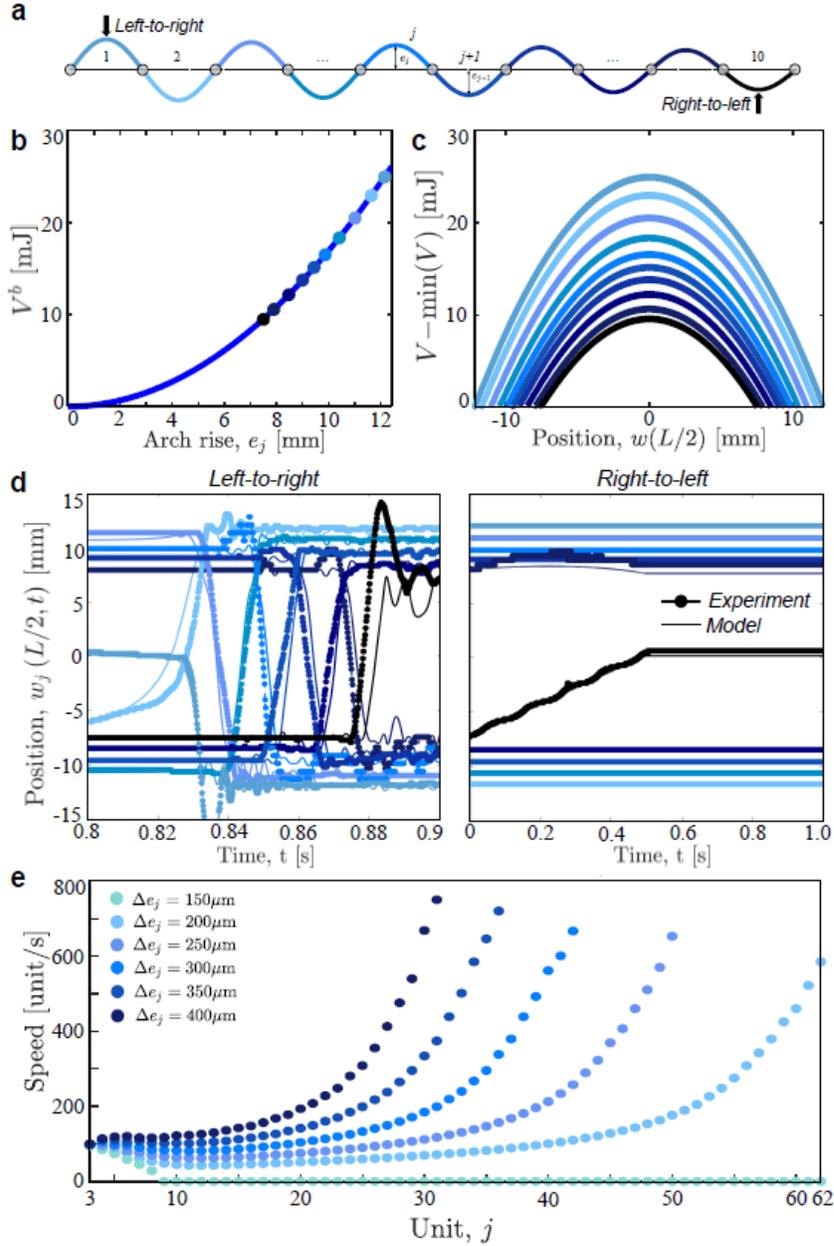

Figure 3: **Symmetric elements - Asymmetric array. a**, Schematic of the system with arches with decreasing rises. **b**, Energy barrier, $V_b$, versus arch rise, $e_j$. **c**, Normalized on-site energy potential, $V - \min(V)$, versus the position of the arch midpoint. **d**, Evolution of the positions of the midpoints of the arches, $w_j(L/2, t)$, for tests in which the indenter acts on the leftmost (left-to-right) and rightmost (right-to-left) arches. Thick- dotted and thin lines correspond to experimental and numerical results, respectively. **e**, Local speed of the transition wave along the chain for different values of $\Delta e_j$ as predicted by our model.

While in Fig. 3d we focus on a specific system with $N = 10$ and $\Delta e_j = e_j - e_{j-1} \sim 500$ μm, we next use our model (which nicely captures the experimental results of Figs. 3d) to systematically investigate the effect of $\Delta e_j$ on the signal propagation in 1D arrays of graded shallow arches. We find that $\Delta e_j$ plays a very important role, as it directly affects the difference in energy barrier between neighboring elements, $\Delta V_j^b$ (see Fig. 3b). More specifically, the numerical results reported in Fig. 3e show that for $\Delta e_j \leq 150$ $\mu$m the velocity of the transition wave (calculated by monitoring the time at which the arches reach the inverted stable configuration - see Fig. S11 for details) monotonically decreases during propagation and eventually vanishes. For such small values of $\Delta e_j$, $\Delta V_j^b$ is not sufficient to overcome the effects of dissipation of the system and a stable wave propagation is not supported. By contrast, for $\Delta e_j > 150$ $\mu$m the difference in energy barriers between consecutive arches is larger than the dissipation upon snapping and the signal propagates through the entire chain. Further, our numerical results indicate that the waves accelerate during propagation. This is because the energy that the arches need to absorb to overcome the energy barrier and snap monotonically decreases along the chain, causing a faster transition rate.

The results of Fig. 3 indicate that graded 1D arrays of shallow arches with symmetric energy landscape can support stable, reversible and unidirectional propagation of transition waves. Next, to achieve additional control on nonreciprocity we introduce elements with asymmetric on-site energy potential. Such asymmetry at the arch level can be easily realized by plastically deforming the metallic shim into the target shape, $e_j \sin(\pi x/L)$. As shown in Figs 4a, the plastic deformation makes the energy minima different. Specifically, for arches with $e_j$ = 12.4 mm our model predicts an energy difference $\Delta V$ = 34.8 mJ between the two stable configurations (Fig 4a). While it has been recently shown that such energy release can be exploited to overcome dissipation and enable propagation of transition waves over long distance (*19*), here we demonstrate that it also provides opportunities to tune nonreciprocity. To this end, we consider a chain comprising seven arches with symmetric on-site potential (elastically deformed arches - see blue arches in Fig. 4b) and three with asymmetric energy profile (plastically deformed arches - see purple arches in Fig. 4b) - all with rise $e_j$ = 12.4 mm. When the arches are arranged as in Fig. 4b (with one plastically deformed arch every two elastically deformed ones), the snapping signal propagates through the entire array both left-to-right and right-to-left (see Fig. 4c). However, the structural asymmetry of the chain leads to different signal propagation in the two directions.

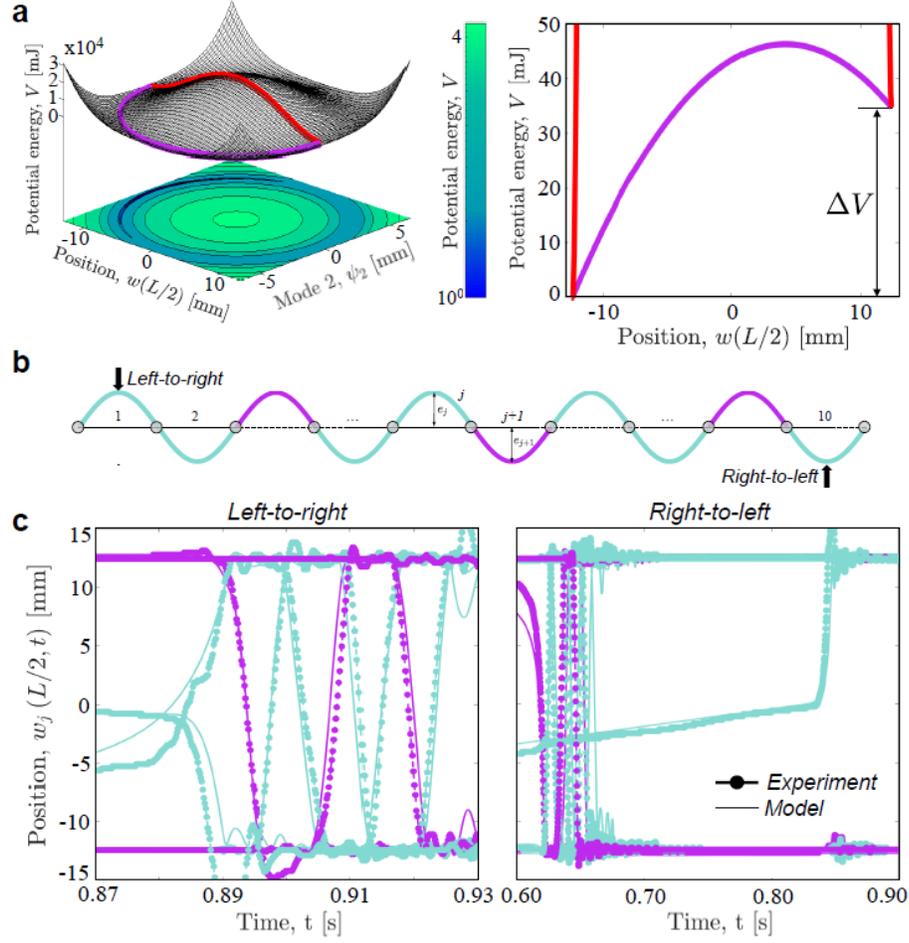

Figure 4: **Symmetric/Asymmetric elements - Asymmetric array. a**, On-site energy potential for a plastically deformed shallow arch. The red line indicates a deformation path along which only the first symmetric mode is activated. The purple line corresponds to the minimum energy path. **b**, Schematic of the system with $N = 10$ shallow arches and one plastically deformed arch arranged every two elastically deformed ones. **c**, Evolution of the positions of the midpoints of the arches, $w_j(L/2, t)$, for tests in which the indenter acts on the leftmost (left-to-right) and rightmost (right-to-left) arches. Thick-dotted and thin lines correspond to experimental and numerical results, respectively.

When the indenter acts on the leftmost unit, the second arch reaches the inverted stable configuration at $t_2^{snap} = 0.89$ s, while the last one snaps at $t_{10}^{snap} = 0.93$ s. By contrast, when the rightmost unit is excited, the pulse is initiated at $t_9^{snap} = 0.62$ s and arrives at the other end of the chain at $t_1^{snap} = 0.66$ s. Interestingly, while for left-to-right propagation the arches snap in sequence (*i.e.* the leftmost arch snaps first and the rightmost one snaps as last), for right-to-left propagation the arch excited by the indenter is the last one to snap at $t_{10}^{snap} = 0.85$ s. Finally, it is important to note that the signal propagation in this system is non-reversible as the plastically deformed arches can only snap from the high energy well to the lower energy well. As such, the chain needs to be manually recharged to support propagation of a new signal (see Fig. S13 and Movie S4).

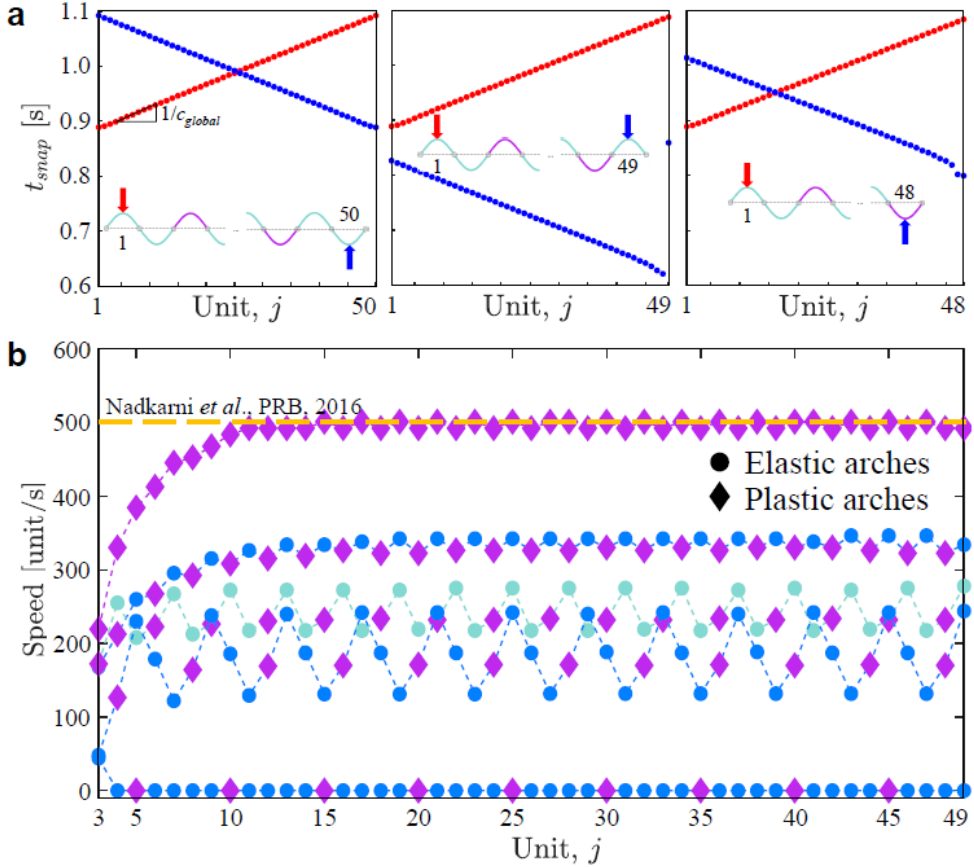

Figure 5: **Tuning nonreciprocity and wave speed. a**, Snapping times for transition waves propagating both left-to-right and right-to-left in chains comprising $N$ = 48, 49 and 50 elastically and plastically deformed arches periodically arranged according to the pattern shown in Fig. 4b. **b**, Local speed of the transition waves along the chain for different patterns of elastically/plastically deformed arches. The yellow dashed line correspond to the analytical prediction of the global speed $c_{global}$ from (42).

Next, to better understand how the global asymmetry affects the nonreciprocity of wave propagation, we numerically investigate the response of chains comprising $N$ = 48, 49 and 50 (Fig. 5a and Fig. S14) elastically and plastically deformed arches periodically arranged according to the pattern shown in Fig. 4b. We find that for all considered chains the pulses propagate at a speed $c_{global}$ ~ 243 unit/s in both directions. However, the time at which the signal is initiated for left-to-right and right-to-left propagation can be programmed by altering the asymmetry of the chain through $N$. More specifically, in a symmetric chain with $N$ = 50 the snapping signal is initiated at the same time for both propagation directions (*i.e.* $t_2^{snap}$ = 0.89 s and $t_{49}^{snap}$ 0.89 s for left-to-right and right-to-left propagation, respectively). Differently, for $N$ = 49 and 48 the system is asymmetric (as there are either one or no elastically deformed arches separating the rightmost plastically deformed one from the right end) and the wave starts at $t_{48}^{snap}$=0.62 s and $t_{47}^{snap}$=0.80 s when the rightmost arch is excited. While asymmetry enables as to tune the time at which the pulses are initiated from the left and right end, control on the speed of the pulses can be achieved by varying the density of plastically deformed elements in the chain. To demonstrate this point, in Fig. 5b we report the numerically predicted velocity for left-to-right propagation in chains with $N$ = 49 plastically and elastically deformed arched arranged according to different periodic patterns. First, we find that stable wave propagation is only possible when the plastically deformed arches are separated by three or less elastically deformed ones. Second, the results indicate that $c_{global}$ monotonically increases with the density of plastically deformed arches and approaches ~ 497 unit/s in the limit of a chain comprising only plastically deformed elements. Note that this maximum speed is very close to the one analytically predicted by balancing the change in the on-site potential energy, dissipation and transported kinetic energy (*42*) (see Fig. S15). Finally, it is interesting to point out that the alternation of elastically and plastically deformed elements leads to pulses with locally modulated speed. This is because the energy released upon snapping by the plastically deformed arches makes the following elastic element to snap faster, whereas the absence of released energy between consecutive elastically deformed ones delays their snapping.

To summarize, we have shown that in 1D multistable systems nonreciprocity and reversability can be programmed independently and easily realized. A reversible diode can be created by assembling elements with symmetric on-site energy potentials but decreasing energy barriers. On the other hand, chains capable of sustaining nonreciprocal transition waves travelling in opposite directions can be realized by alternating arches with symmetric and asymmetric energy wells. While the dynamic response of the reversible diode is controlled by the difference in rise between the arches, the behavior of the nonreciprocal chain can be tuned by varying the arrangement of the symmetric and asymmetric elements and is negligibly affected by their rise (see Fig. S16). Further, all the considered systems are input-independent, as the speed of the supported waves is insensitive to the loading rate at which the indenter moves the first arch (see Fig. S17). Although in this study we verified the concept for 1D chains, our findings can be easily generalized to 2D and 3D network of arches to realize passive smart systems with unprecedented selective signal transmission and wave guidance capabilities.

K.B. acknowledges support from NSF grants EFMA-1741685 and DMR-2011754 and Army Research Office Grant W911NF-17-1-0147. E.T. acknowledges support from University of Maryland, College Park - startup package. G.L. and E.T. acknowledges John Hutchinson and Lakshminarayanan Mahadevan for interesting conversations and helpful suggestions.

# Supporting Information for
# *Programming nonreciprocity and reversibility in multistable mechanical metamaterials*

## S1 Fabrication

The structures considered in this study comprise a 1D array of shallow arches made out of metallic beams and connected via rotating hinges. All beams have width $b = 10$ mm, length $l \in [103.1, 105.0]$ mm and are made of spring steel shims (McMaster-Carr product ID: 9014K611) with thickness $h = 0.3048$ mm and Young's modulus $E = 170$ GPa, while all hinges are realized using Lego components. Specifically, as shown in Fig. S1, the hinges comprise the following components: (*i*) a round brick 2 × 2 with axle hole (LEGO part 4249139); (*ii*) an axle (LEGO part 3705); (*iii*) axle and pin connectors angled at 180 degrees (LEGO part 32034). Note that in order to ensure a tight connection between the beam and the hinges, slits of 3 mm length and 0.33 mm width are cut into both arms using a vertical knee drilling milling machine with slitting saw (Vectrax); (*iv*) two bushes (LEGO part 32123); and (*v*) a round brick 4 × 4 with a center axle hole (LEGO part 4211097). Note that for a structure with $N$ arches we use $N + 1$ of these hinges with their 4 × 4 round brick connected and glued (using Krazy Glue *All purposes*) to a Lego plate (LEGO part 91405). The axels of the hinges (LEGO part 3705) are located at a distance $L = 120$ mm from each other.

In the remaining part of this Section we first describe how we fabricate chains comprising only *elastically deformed* shallow arches and then report our manufacturing process to realize *plastically deformed* arches that are introduced into the chain in selected locations.

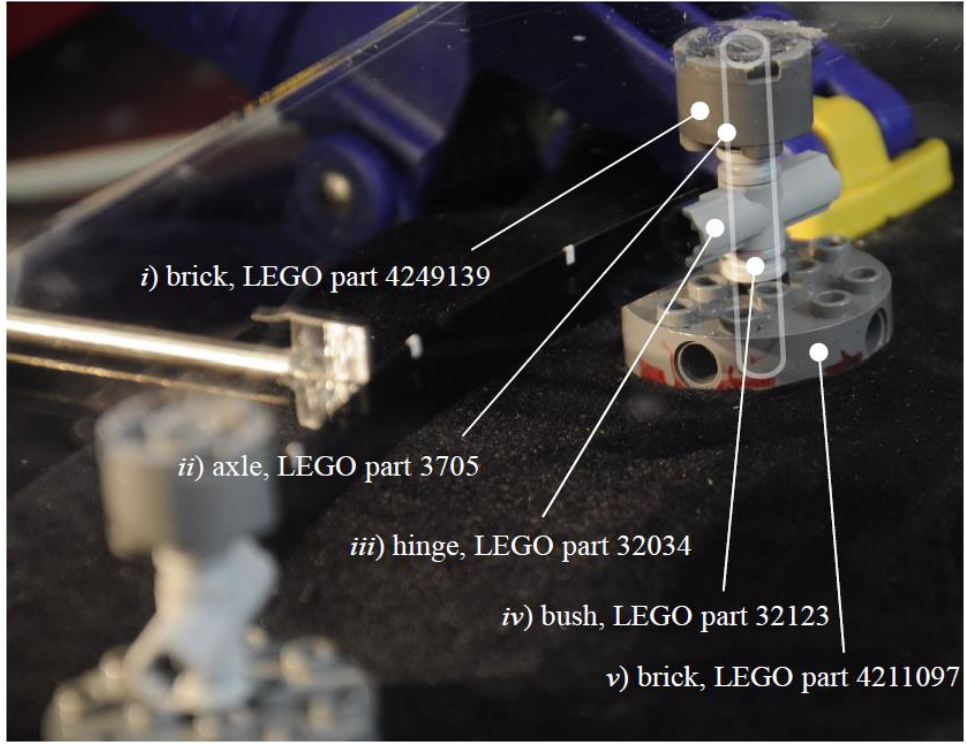

Figure S1: **Hinges used in the experimental campaign.** Picture of the hinge used to build our arches. The parts used to fabricate them are highlighted.

## S1.1   Chains comprising only *elastically deformed* shallow arches

As shown in Fig. S2, a chain comprising $N$ elastically deformed arches is fabricated using the following 3 steps:

**Step 1:** $N$ strips of length $l$ (with $l \in [103.1, 105.0]$ mm) and width $b = 10$ mm are cut out of the steel shim by using a shearing tool. The strips are visually inspected to make sure they do not have any residual bending induced by the cutting procedure.

**Step 2:** Both ends of all steel strips are inserted into the cuts of the Lego connectors to form a 1D chain. Glue (Krazy Glue *All purposes*) is applied to prevent any sliding.

**Step 3:** An axial force is sequentially applied to all the strips to buckle them and form an array of *N* arches with end-to-end distance $L_{TOT} = N\,L$. Specifically, the axles (LEGO part 3705) are slid into the hinges connected to the arches (LEGO part 32034) and LEGO bushes (LEGO part 32123) are added on the axles to prevent other movements rather than rotation of the hinges. Lastly, an acrylic plate with LEGO round bricks (LEGO part 4249139) glued on it is fixed on the top to prevent bending of the axles. The array of arches is now ready to be tested.

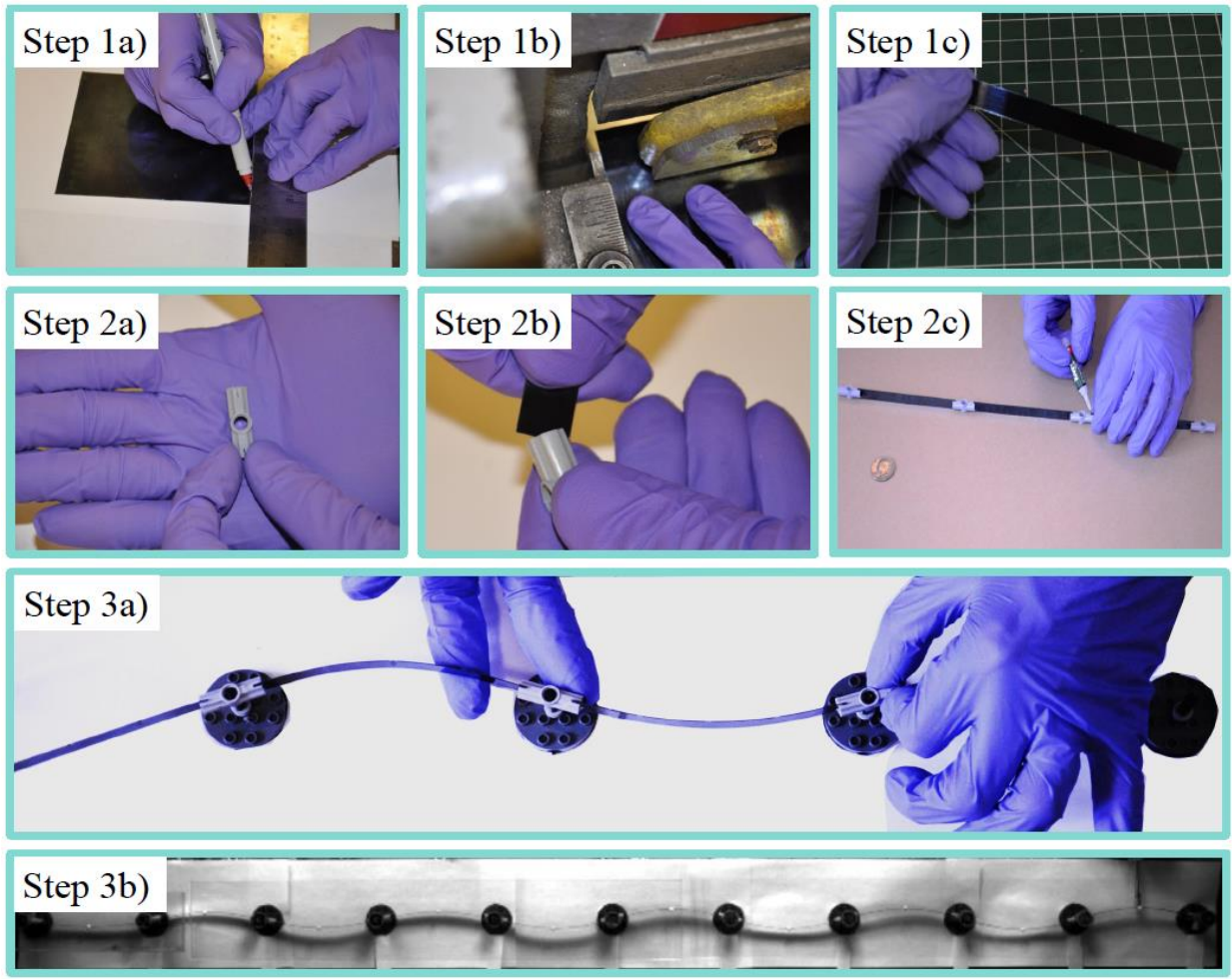

Figure S2: **Fabrication.** Fabrication steps to manufacture a chain comprising only *elastically deformed* shallow arches.

## S1.2 Chains comprising *elastically deformed* and *plastically deformed* shallow arches

As part of this study, we also fabricated structures comprising $N_{el}$ elastically deformed arches and $N_{pl}$ plastically deformed shallow arches introduced in selected locations. As shown in Fig. S3, these structures are fabricated using the following 5 steps:

**Step 1:** $N_{el}$ strips of length $l$ =105.0 mm and width $b$ = 10 mm are cut out of the steel shim by using a shearing tool. The strips are visually inspected to make sure they do not have any residual bending induced by the cutting procedure.

**Step 2:** $N_{pl}$ strips of length $l$ =105.0 mm and width $b$ = 10 mm are cut with a shearing tool out of the steel shim. The strips are visually inspected to make sure they do not have any residual bending induced by the cutting procedure.

**Step 3:** Iron cylinders of diameter 34 and 25 mm are used to plastically deform the $N_{pl}$ steel beams into a sinusoidal-like shape.

**Step 4:** The shape of the $N_{pl}$ plastically deformed arch is visually compared with that of the target sinusoidal profile (which is laser cut out of an acrylic sheet with thickness 12.7 mm) to make sure the obtained shape is close enough to the desired one. In the unlikely event the arch has a very different shape with respect the benchmark, we either repeat Step 3 or start over from Step 2.

**Step 5:** Both ends of the $N_{el}$ steel strips and the $N_{pl}$ plastically deformed arches are inserted into the cuts of the LEGO connectors to form a 1D chain (with the plastically deformed arches arranged in the desired location). Glue (Krazy Glue *All purposes*) is applied to prevent any sliding. The gluing is repeated on all the steel strips that are assembled together to form our array of arches.

**Step 6:** The $N_{pl}$ plastically deformed arches are directly connected to the LEGO supports. An axial force is sequentially applied to the $N_{el}$ strips to buckle them and form arches that are then connected to the Lego supports. Specifically, the axles (LEGO part 3705) are slid into the hinges connected to the arches (LEGO part 32034) and LEGO bushes (LEGO part 32123) are added on the axles to prevent other movements rather than rotation of the hinges. Lastly, an acrylic plate with LEGO round bricks (LEGO part 4249139) glued on it is fixed on the top to prevent bending of the axles. The array of arches is now ready to be tested.

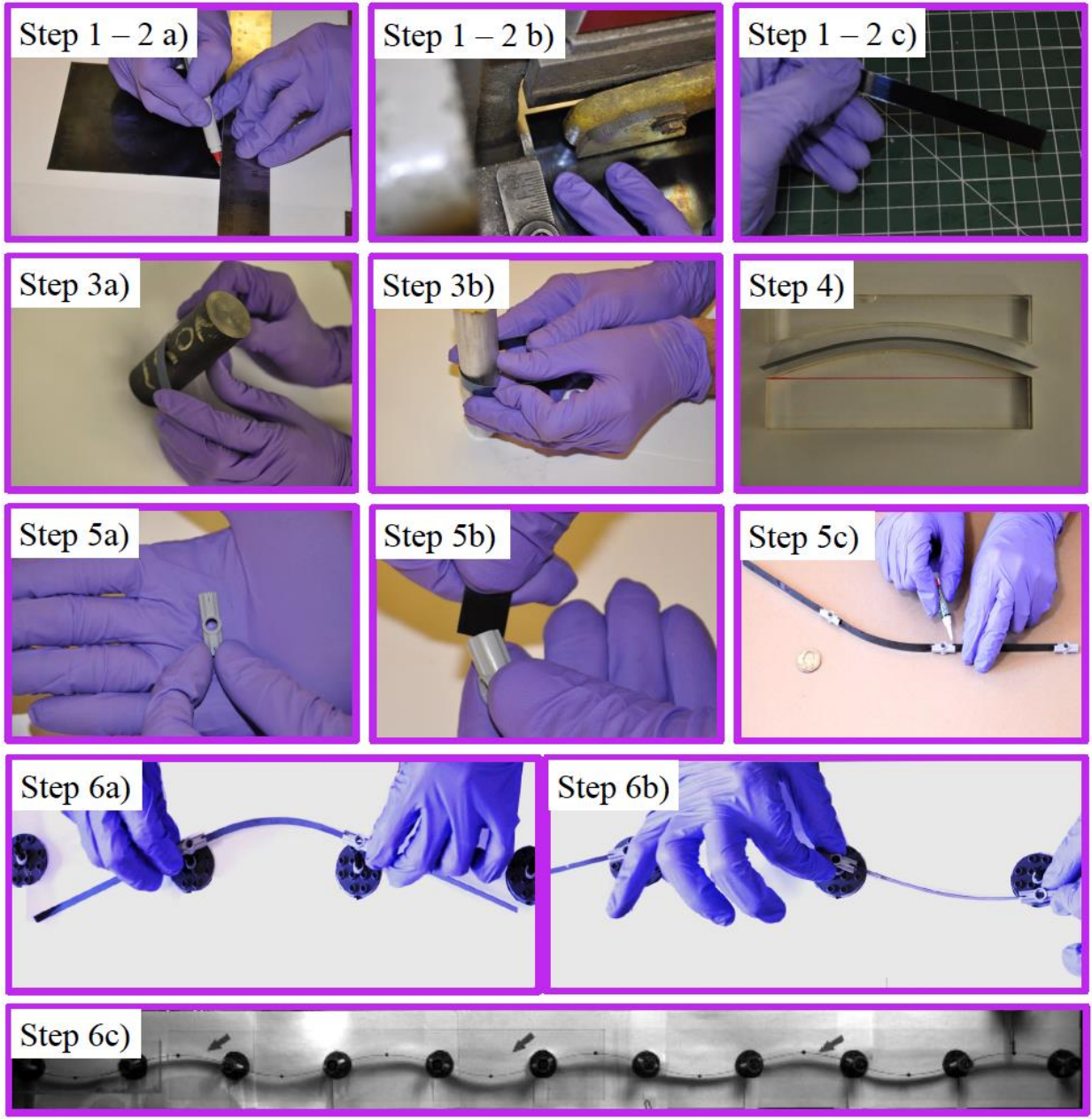

Figure S3: **Fabrication.** Fabrication steps to manufacture a chain comprising *elastically deformed* and *plastically deformed* shallow arches.

# S2 Testing

In all our tests we used an indenter (see inset in Fig. S4) to push the central part of the first arch in the chain at a constant velocity of 15 mm/s (via a motorized translation stage - LTS300, Thorlabs). During our tests the reaction force is measured using a 10 lb load cell (LSB200 Miniature S-Beam Jr. Load cell, FUTEK Advanced Sensor Technology, Inc.). Moreover, a highspeed camera (Photron Mini Series) is mounted above the testing area to track the displacements of the markers positioned at $L/2$ of each arch comprising our specimen. The camera records the entire experiment from the contact between the arch and the indenter up to the snapping of the entire array at a rate of 6400 fps.

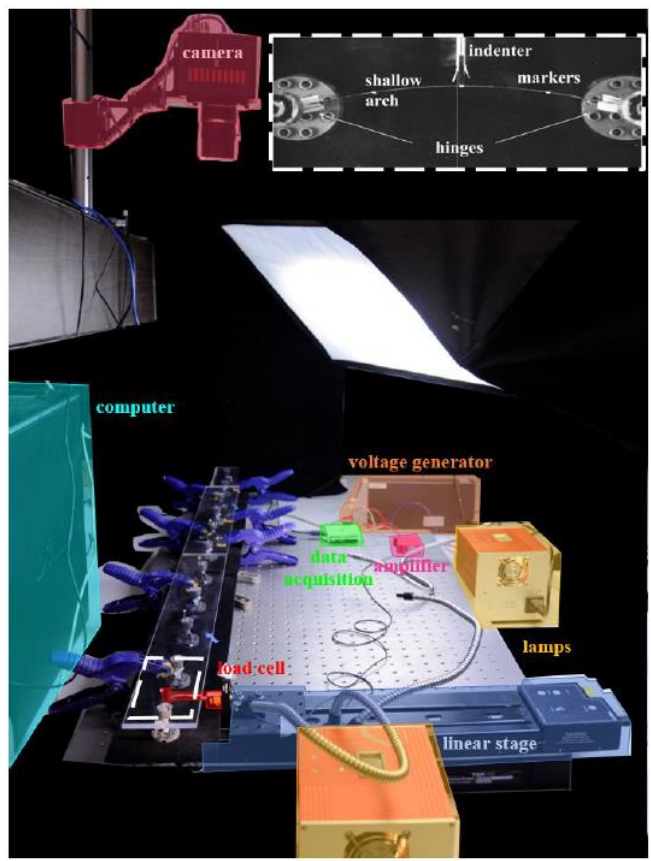

Figure S4: **Experimental setup.** Components of the experimental setup used to test our array of arches

## S3 Mathematical model

To get a better understanding of the dynamic response of our system, we establish a numerical model. As in our experiments, we consider a 1D array comprising $N$ shallow arches connected via rotating hinges, see Fig. S5. Focusing on the $j$-th arch ($j \in [1, N]$), we describe its initial shape as

$$w_j(x_j, t=0) = e_j \sin\left(\frac{\pi x_j}{L}\right) \tag{S1}$$

where $x_j \in [0, L]$ ($L$ denoting the span of the arches) and the rise $e_j$ is positive if the arch is curved upwards and negative if the arch is curved downwards. We then apply a monotonically increasing displacement $d(t)$ ($d(t) < 0$ when pushing downwards) to the midpoint of the first arch in the array (note that the equations can be very easily adjusted to account for the loading of any other arch in the array) and solve for the time-dependent transverse profile of the $j$-th arch, $w_j(x_j, t)$ (which is defined positive for positive values of $z$). Towards this end, we use Kirchhoff plate theory (1) to describe the behavior of the individual arches (2–8), so that their potential energy is given by

$$V_j = \frac{1}{2}EI \int_0^L \left(\frac{\partial^2 w_j}{\partial x_j^2} - \frac{d^2 w_{0j}}{dx_j^2}\right)^2 dx_j - \frac{1}{2}P_j \int_0^L \left(\frac{\partial w_j}{\partial x_j}\right)^2 dx_j + \frac{EA}{8L}\left[\int_0^L \left(\frac{\partial w_j}{\partial x_j}\right)^2 - \left(\frac{dw_{0j}}{dx_j}\right)^2 dx_j\right]^2 \tag{S2}$$

where $w_{0j}$ is the initial unstressed position of the midsurface of the $j$-th arch, $P_j$ is the axial force applied to the $j$-th arch to elastically buckle it. Moreover, $A$ and $I$ are the area and moment of inertia of the cross section, $E$ is the Young's modulus of the material. Further, we impose continuity of rotations between neighboring elements (9, 10)

$$\left.\frac{\partial w_{j-1}}{\partial x_{j-1}}\right|_{x_{j-1}=L} = \left.\frac{\partial w_j}{\partial x_j}\right|_{x_j=0}. \tag{S3}$$

Importantly the constraints described by Eq. (S3) introduce concentrated moments, $M_{Lj}$ and $M_{Rj}$, at both ends of the $j$-th arch

$$M_{R_j} = M_{Rj}(t)\delta'(x_j - L), \quad M_{L_j} = M_{Lj}(t)\delta'(x_j), \tag{S4}$$

$\delta$ being the Dirac delta function, that satisfy

$$M_{R_{j-1}} = -M_{L_j}, \quad M_{R_j} = -M_{L_{j+1}}. \tag{S5}$$

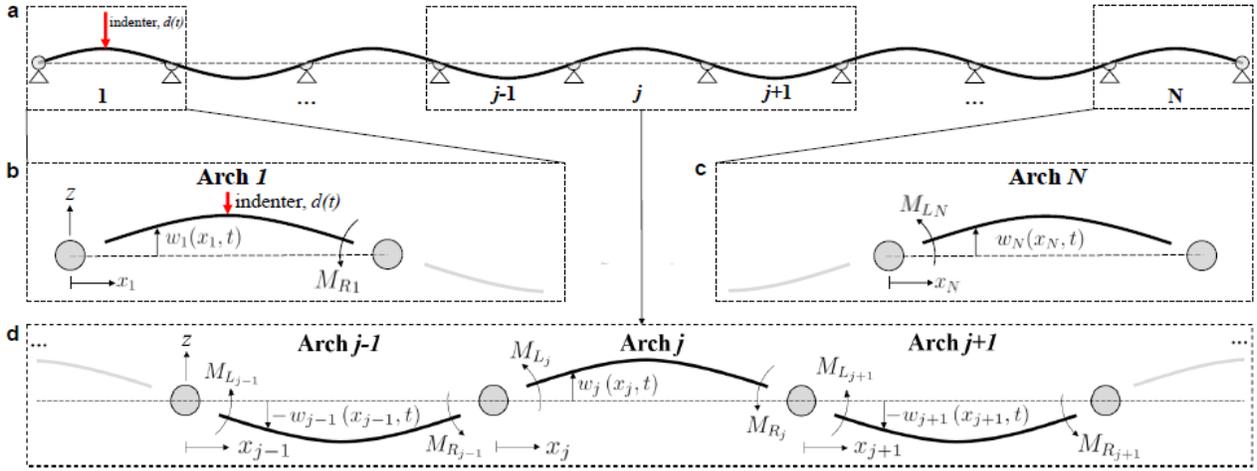

Figure S5: **Schematic. a**, Schematic of a 1D array comprising $N$ shallow arches. **b**, Schematic of the arch where the indenter is applied. **c**, Schematic of the last arch in the chain. **d**, Schematic of a central portion of the chain where arches interact with neighboring elements both on the left and on the right hand side.

It follows that the response of an array comprising $N$ arches can be described as

$$\rho A \frac{\partial^2 w_j}{\partial t^2} + \beta \frac{\partial w_j}{\partial t} + EI \left( \frac{\partial^4 w_j}{\partial x_j^4} - \frac{d^4 w_{0j}}{dx_j^4} \right) + p_j \frac{\partial^2 w_j}{\partial x_j^2} + Q_1 + M_{Rj} = 0 \tag{S6a}$$

for $j = 1$

$$\rho A \frac{\partial^2 w_j}{\partial t^2} + \beta \frac{\partial w_j}{\partial t} + EI \left( \frac{\partial^4 w_j}{\partial x_j^4} - \frac{d^4 w_{0j}}{dx_j^4} \right) + p_j \frac{\partial^2 w_j}{\partial x_j^2} + M_{Lj} + M_{Rj} = 0 \tag{S6b}$$

for $j = 2, .., N-1$

$$\rho A \frac{\partial^2 w_j}{\partial t^2} + \beta \frac{\partial w_j}{\partial t} + EI \left( \frac{\partial^4 w_j}{\partial x_j^4} - \frac{d^4 w_{0j}}{dx_j^4} \right) + p_j \frac{\partial^2 w_j}{\partial x_j^2} + M_{Lj} = 0 \tag{S6c}$$

for $j = N$

where $p_j = p_j(t)$ is the midplane force produced by the stretching of the middle surface of the $j$-th arch. Moreover, $\rho$ and $\beta$ are the volumetric density and the viscous damping coefficient, respectively. Further, $Q_1$ denotes the force measured at the midpoint of the first arch,

$$Q_1 = Q_1(t) \, \delta \left( x_1 - \frac{L}{2} \right) \tag{S7}$$

where $\delta$ is the Dirac delta function.

Additionally, since we are considering an indenter that controls the displacement of the first arch, we impose

$$w_1 \left( \frac{L}{2}, t \right) = e_1 + d(t), \tag{S8}$$

which holds true until the first arch is in contact with the indenter. Note that if the $j$-th arch is elastically deformed, the midplane force is given by (2, 7, 11)

$$p(t) = \left[ \frac{EA}{L_{0j}} (L_{0j} - L) - \frac{EA}{2L} \int_0^L \left( \frac{\partial w_j}{\partial x_j} \right)^2 dx_j \right], \tag{S9}$$

and the initial unstressed position of the midsurface is

$$w_{0j} = 0, \tag{S10}$$

where $L_{0j}$ denotes the length of the $j$-th beam in its undeformed configuration.

Differently, if the *j*-th arch is plastically deformed, the midplane force is given by (2,4,12)

$$p(t) = -\frac{EA}{2L}\int_0^L \left[\left(\frac{\partial w_j}{\partial x_j}\right)^2 - \left(\frac{dw_{0j}}{dx_j}\right)^2\right] dx_j, \quad (S11)$$

and the initial unstressed position of the midsurface is

$$w_{0j}(x_j) = e_j \sin\left(\frac{\pi x_j}{L}\right). \quad (S12)$$

Importantly, for our system the deformed shape $w_j(x_j, t)$ can be expressed as a series of sine functions (2)

$$w_j(x_j, t) = w_{0j}(x_j) + \sum_{n=1}^{N_t} \psi_{nj}(t) \sin\left(\frac{n\pi x_j}{L}\right). \quad (S13)$$

If the *j*-th arch is an elastically deformed one, substitution Eqs. (S9), (S10), and (S13) into Eqs. (S6), multiplication of all terms by sin (*mπxj/L*) (*m* being an integer, *m* = 1, .., $N_t$) and integration with respect to $x_j$ from 0 to *L* yields

$$\frac{\rho A}{2}\ddot{\psi}_{nj} + \frac{\beta}{2}\dot{\psi}_{nj} + \frac{EI}{2}\left(\frac{n\pi}{L}\right)^4 \psi_{nj} - \frac{EAL}{2}\left(\frac{n\pi}{L}\right)^2 \left(\frac{L_{0j}-L}{L} - \frac{1}{4}\sum_{k=1}^{N_t}\left(\frac{k\pi}{L}\right)^2 \psi_{kj}^2\right)\psi_{nj} + q_n + m_{R_{nj}} = 0$$

$$\text{for } j=1, \quad n=1,...,N_t$$
$$(S14a)$$

$$\frac{\rho A}{2}\ddot{\psi}_{nj} + \frac{\beta}{2}\dot{\psi}_{nj} + \frac{EI}{2}\left(\frac{n\pi}{L}\right)^4 \psi_{nj} - \frac{EAL}{2}\left(\frac{n\pi}{L}\right)^2 \left(\frac{L_{0j}-L}{L} - \frac{1}{4}\sum_{k=1}^{N_t}\left(\frac{k\pi}{L}\right)^2 \psi_{kj}^2\right)\psi_{nj} + m_{R_{nj}} + m_{L_{nj}} = 0$$

$$\text{for } j \in [2,...,N-1], \quad n=1,...,N_t$$
$$(S14b)$$

$$\frac{\rho A}{2}\ddot{\psi}_{nj} + \frac{\beta}{2}\dot{\psi}_{nj} + \frac{EI}{2}\left(\frac{n\pi}{L}\right)^4 \psi_{nj} - \frac{EAL}{2}\left(\frac{n\pi}{L}\right)^2 \left(\frac{L_{0j}-L}{L} - \frac{1}{4}\sum_{k=1}^{N_t}\left(\frac{k\pi}{L}\right)^2 \psi_{kj}^2\right)\psi_{nj} + m_{L_{nj}} = 0$$

$$\text{for } j=N, \quad n=1,...,N_t$$
$$(S14c)$$

where

$$q_n = Q \sin\left(\frac{n\pi}{2}\right), \quad m_{R_{nj}} = -M_{Rj}\left(\frac{n\pi}{L}\right)\cos(n\pi), \quad m_{L_{nj}} = -M_{Lj}\left(\frac{n\pi}{L}\right). \quad (S15)$$

Moreover, by substituting Eqs. (S10) and (S13) into Eq. (S8) we obtain

$$-d(t) - e_1 + \sum_{n=1}^{\|(N_t+1)/2\|} -(-1)^n \psi_{2n-1} = 0. \quad (S16)$$

Differently, if the $j$-th arch is a plastically deformed one, substitution Eqs. (S11), (S12), and (S13) into Eqs. (S6b)-(S6c), multiplication of all terms by sin $(m\pi x_j/L)$ ($m$ being an integer, $m = 1, .., N_t$) and integration with respect to $x_j$ from 0 to $L$ yields

$$\frac{\rho A}{2}\ddot{\psi}_{nj} + \frac{\beta}{2}\dot{\psi}_{nj} + \frac{EI}{2}\left(\frac{\pi}{L}\right)^4 \psi_{nj} + \frac{EA}{4L}\left(\frac{\pi}{L}\right)^2\left(\frac{2e_j\pi}{L}\psi_{nj} + \sum_{k=1}^{N_t}\left(\frac{k\pi}{L}\right)^2 \psi_{kj}^2\right)(\psi_{nj} + e_j) + m_{R_{nj}} + m_{L_{nj}} = 0$$

$$\text{for } j \in [2, ..., N-1], \quad n = 1$$

$$\frac{\rho A}{2}\ddot{\psi}_{nj} + \frac{\beta}{2}\dot{\psi}_{nj} + \frac{EI}{2}\left(\frac{n\pi}{L}\right)^4 \psi_{nj} + \frac{EA}{4L}\left(\frac{n\pi}{L}\right)^2\left(\frac{2e_j\pi}{L}\psi_{nj} + \sum_{k=1}^{N_t}\left(\frac{k\pi}{L}\right)^2 \psi_{kj}^2\right)\psi_{nj} + m_{R_{nj}} + m_{L_{nj}} = 0$$

$$\text{for } j \in [2, ..., N-1], \quad n = 2, ..., N_t$$

$$(S17a)$$

$$\frac{\rho A}{2}\ddot{\psi}_{nj} + \frac{\beta}{2}\dot{\psi}_{nj} + \frac{EI}{2}\left(\frac{\pi}{L}\right)^4 \psi_{nj} + \frac{EA}{4L}\left(\frac{\pi}{L}\right)^2\left(\frac{2e_j\pi}{L}\psi_{nj} + \sum_{k=1}^{N_t}\left(\frac{k\pi}{L}\right)^2 \psi_{kj}^2\right)(\psi_{nj} + e_j) + m_{L_{nj}} = 0$$

$$\text{for } j = N, \quad n = 1$$

$$\frac{\rho A}{2}\ddot{\psi}_{nj} + \frac{\beta}{2}\dot{\psi}_{nj} + \frac{EI}{2}\left(\frac{n\pi}{L}\right)^4 \psi_{nj} + \frac{EA}{4L}\left(\frac{n\pi}{L}\right)^2\left(\frac{2e_j\pi}{L}\psi_{nj} + \sum_{k=1}^{N_t}\left(\frac{k\pi}{L}\right)^2 \psi_{kj}^2\right)\psi_{nj} + m_{L_{nj}} = 0$$

$$\text{for } j = N, \quad n = 2, ..., N_t$$

$$(S17b)$$

where $m_{Rnj}$, and $m_{Lnj}$ are given by Eq. (S15). Note that, since in all our analyses the first arch in the chain is always an elastically deformed one, in Eqs. (S17) we do not include the case $j = 1$.

Eqs. (S3), (S5), (S14), (S16), and (S17) are solved by numerical integration using the Runge-Kutta method (via the ODE45 function in Matlab) to obtain $\psi_{nj}(t)$. Note that in all our analyses we use $\rho$ = 7850 $kg/m^3$, $A$ = 3.05 $mm^2$, $E$ = 170 $GPa$, $I$ = 0.02 $mm^4$ (all values that are measured), $\beta$ = 6.71 · $10^{-7}$ $kg/(m \cdot s)$ for chains comprising only elastically deformed shallow arches and $\beta$ = 1.41 · 10$^{-6}$ *kg/(m · s)* for chains comprising both elastically and plastically deformed shallow arches (note that these values are chosen to better capture the response observed in our tests).

At this point it is important to emphasize that Eqs. (S8) and (S14a) describe the behavior of the first arch when this is in contact with the indenter (*i.e.* until $|Q_1(t)|$ > 0) (*2*). When $|Q_1(t)|$ = 0 the first arch leaves the indenter, Eq. (S8) does not hold true anymore and Eq. (S14a) simplify to

$$\frac{\rho A}{2}\ddot{\psi}_{nj} + \frac{\beta}{2}\dot{\psi}_{nj} + \frac{EI}{2}\left(\frac{n\pi}{L}\right)^4 \psi_{nj} - \frac{EAL}{2}\left(\frac{n\pi}{L}\right)^2 \left(\frac{L_{0j} - L}{L} - \frac{1}{4}\sum_{k=1}^{N_t}\left(\frac{k\pi}{L}\right)^2 \psi_{kj}^2\right)\psi_{nj} + m_{R_{nj}} = 0$$
$$\text{for} \quad j = 1, \quad n = 1, ..., N_t$$
(S18a)

The system comprising Eqs. (S3), (S5), (S18), Eqs. (S14b)-(S14c) for elastically deformed arches and Eqs. (S17) for plastically deformed arches is again numerically integrated using the Runge-Kutta method (via the ODE45 function in Matlab) with initial conditions (positions and velocities) given by Eqs. (S16) and Eqs. (S14) or Eqs. (S17) for elastically and plastically deformed arches, respectively.

To determine the number $N_t$ of modes required for our model to accurately capture the response of our chains, we first focus on an array comprising $N$ = 3 elastically deformed shallow arches with $e_j$ = 12.4 mm. When considering $N_t$ = 5 modes, we find that both $\psi_4$ and $\psi_5$ are negligible with respect the first three modes (*i.e.* $\psi_j$ with $j$ = 1, 2, 3) during the entire simulation (see Fig. S6). As such, these results indicate that $N_t$ = 3 modes are sufficient to capture the response of chains comprising elastically deformed arches. Next, we focus on an array comprising 3 elastically deformed arches and a plastically deformed one. As shown in Fig. S7, also for this case $\psi_4$ and $\psi_5$ are negligible during the entire simulation. Therefore, $N_t$ = 3 modes are sufficient to capture also the response of chains comprising both elastically and plastically deformed arches.

Finally, we note that substitution of Eq. (S13) into Eq. (S2) yields

$$V_j = \frac{\pi^4 EI}{4L^3} \sum_{n=1}^{N_t} n^4 \psi_{nj}^2 - EA\frac{\pi^2}{4L}\frac{L_0 - L}{L_0} \sum_{n=1}^{N_t} n^2 \psi_{nj}^2 + \frac{\pi^4 EA}{32L^3}\left[\sum_{n=1}^{N_t} n^2 \psi_{nj}^2\right]^2 \quad (S19)$$

for elastically deformed arches and

$$V_j = \frac{\pi^4 EI}{4L^3} \sum_{n=1}^{N_t} n^4 \psi_{nj}^2 + \frac{\pi^4 EA}{32L^3}\left[2w_{0j}\psi_{1j} + \sum_{n=1}^{N_t} n^2 \psi_{nj}^2\right]^2 \quad (S20)$$

for plastically deformed ones. Eqs. (S19) and (S20) with $N_t = 3$ are used to calculate the energy landscapes for the elastically and plastically deformed arches shown in Figs. 2f and 4a of the main text.

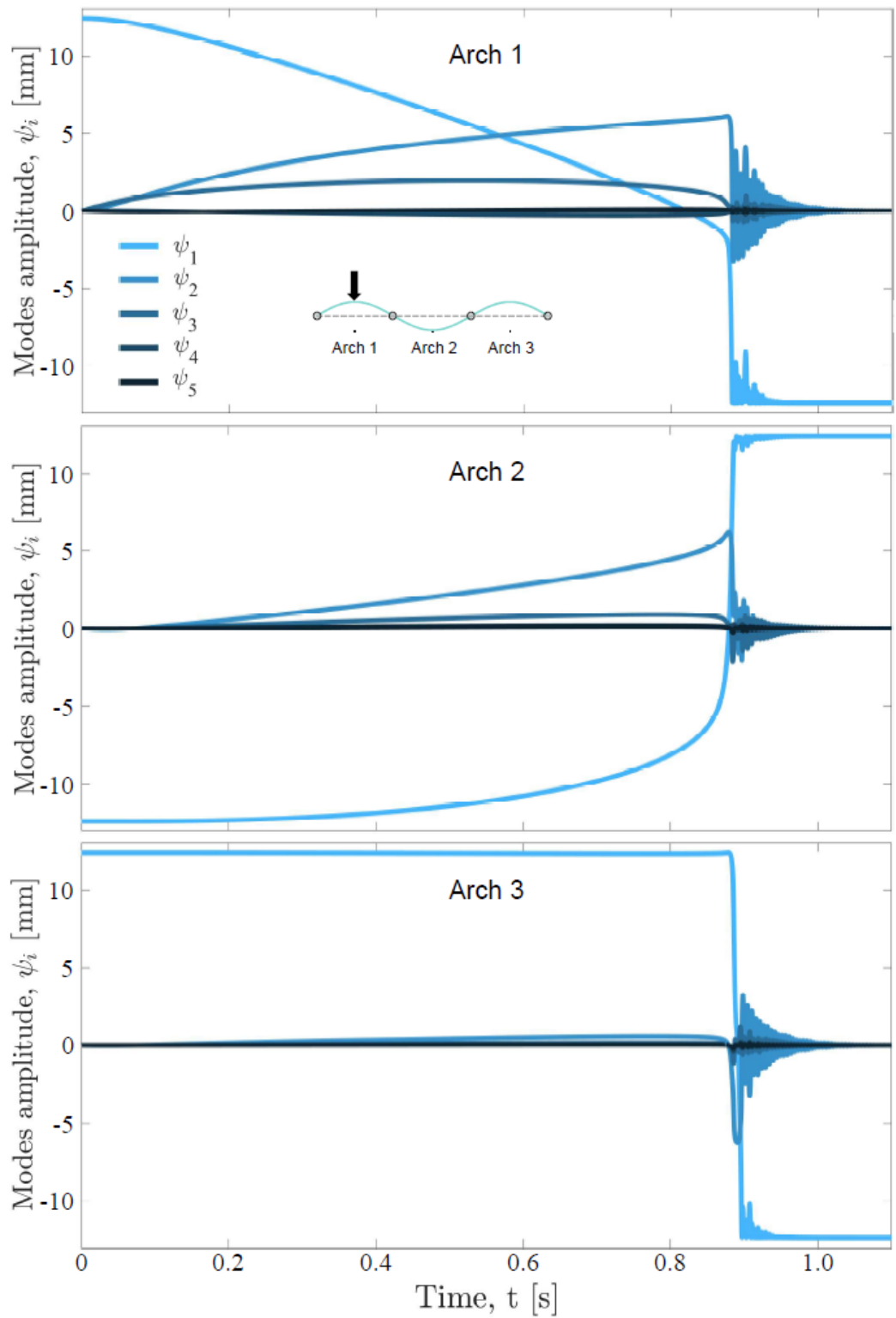

Figure S6: **Effect of $N_t$ on the response of an array comprising only elastic arches.**

Convergence analysis for a 1D array with $N = 3$ elastically deformed shallow arches. For all the three arches both the fourth, $\psi_4$, and the fifth, $\psi_5$, modes are negligible.

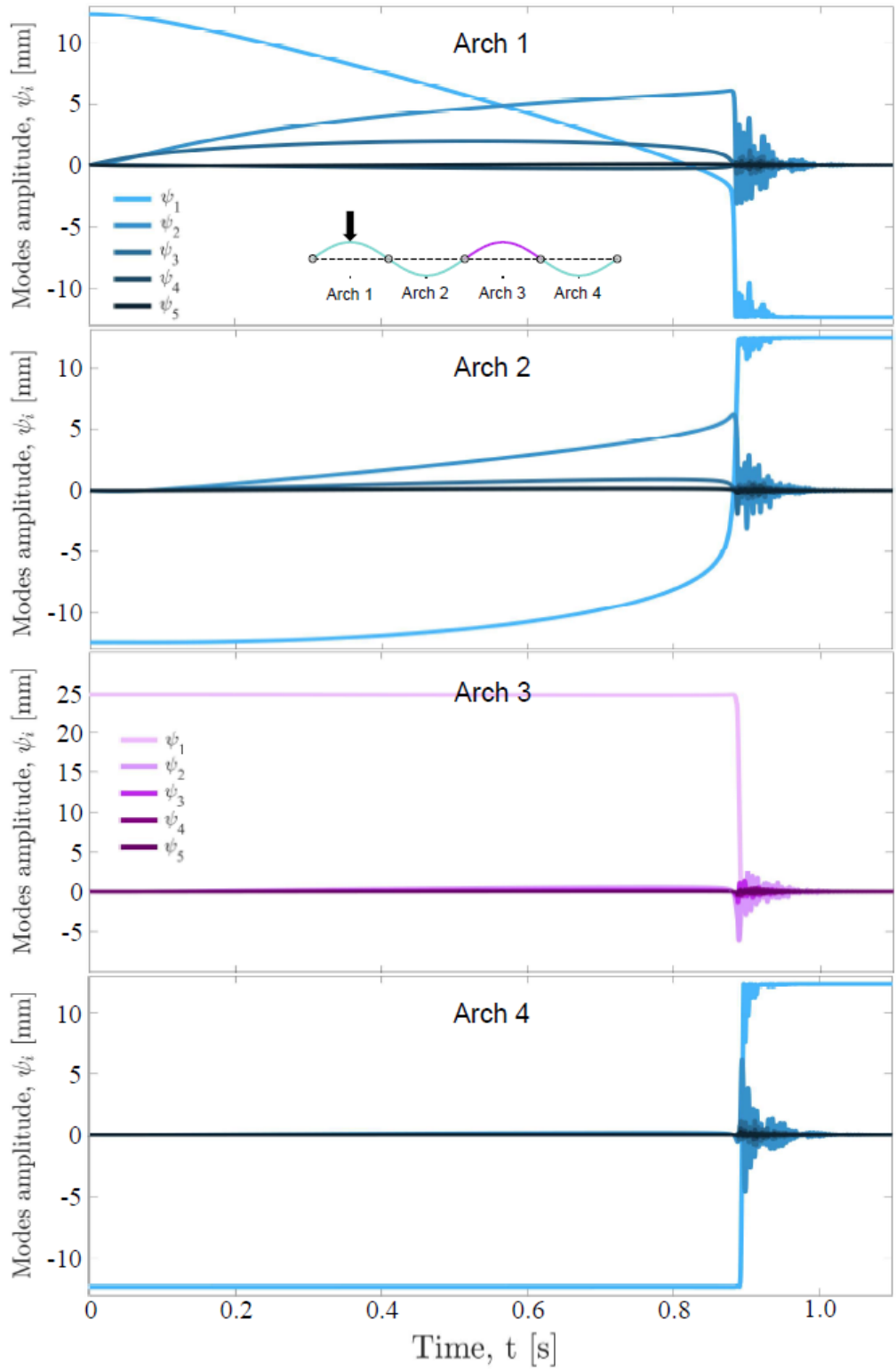

Figure S7: **Effect of $N_t$ on the response of an array comprising elastic and plastic arches.**

Convergence analysis for a 1D array with 3 elastically deformed shallow arches and one plastically deformed one. For all the three arches both the fourth, $\psi_4$, and the fifth, $\psi_5$, modes are negligible.

## S4 Additional results

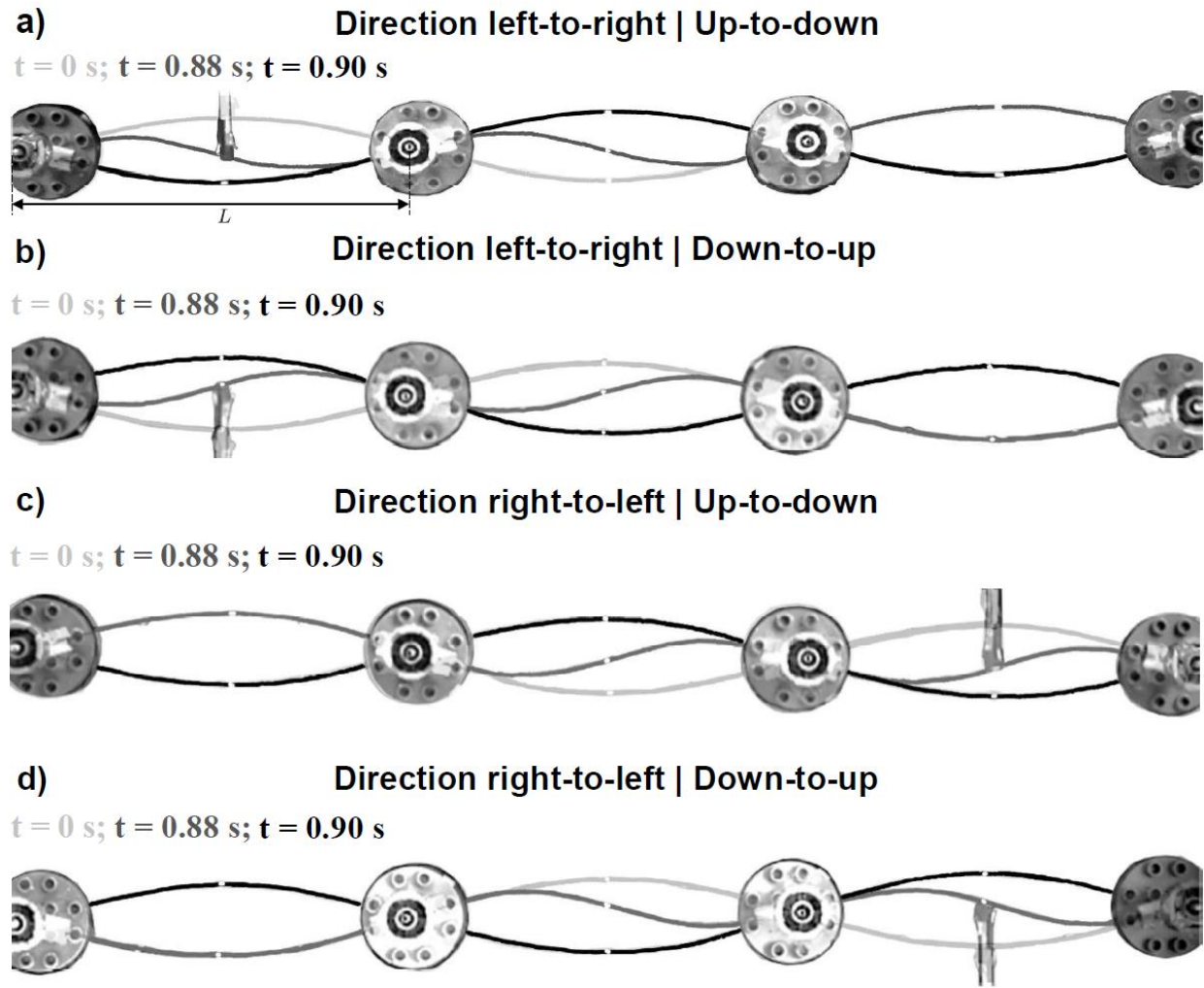

Figure S8: **Symmetric elements – Symmetric array.** Chain comprising three identical arches with rise $e_j = 12.4$ mm ($j = 1,2,3$) and symmetric energy wells. **a**, Signal propagation left-to-right by exciting the leftmost arch from up-to-down. **b**, Signal propagation left-to-right by exciting the leftmost arch from down-to-up. **c**, Signal propagation right-to-left by exciting the rightmost arch from up-to-down. **d**, Signal propagation right-to-left by exciting the rightmost arch from down-to-up.

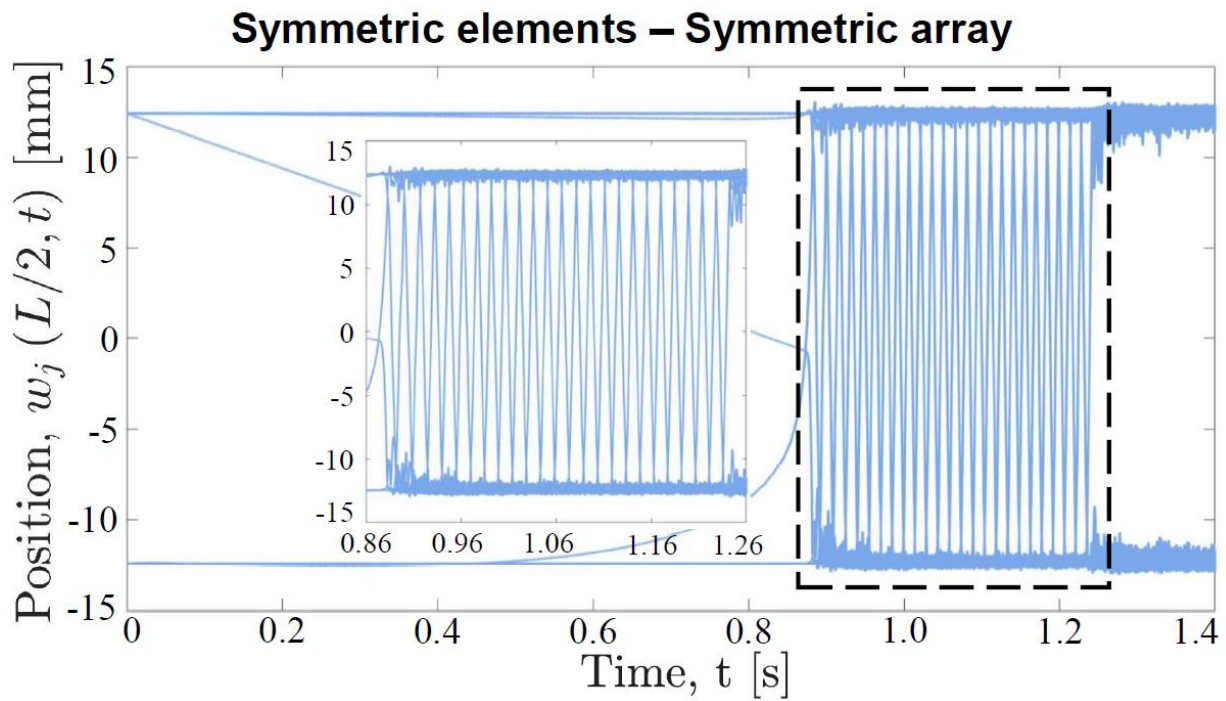

Figure S9: **Symmetric elements – Symmetric array: response of an ideal chain with no damping.** Numerical simulation of an array comprising $N$ = 50 elastically deformed shallow arches all with rise $e_j = 12.4$ mm in the absence of damping (i.e. $\beta = 0$). An ideal system with all identical elastic elements and no damping can potentially sustain a transition wave over arbitrary distances.

Table S1: Measured rises of the elastically deformed shallow arches in the chain considered in Fig. 3 of the main text.

|         | $w_j(L/2, 0)$ [mm] |
|---------|--------------------|
| Arch 1  | 12.13              |
| Arch 2  | 11.64              |
| Arch 3  | 11.00              |
| Arch 4  | 10.63              |
| Arch 5  | 9.88               |
| Arch 6  | 9.59               |
| Arch 7  | 9.03               |
| Arch 8  | 8.49               |
| Arch 9  | 7.92               |
| Arch 10 | 7.51               |

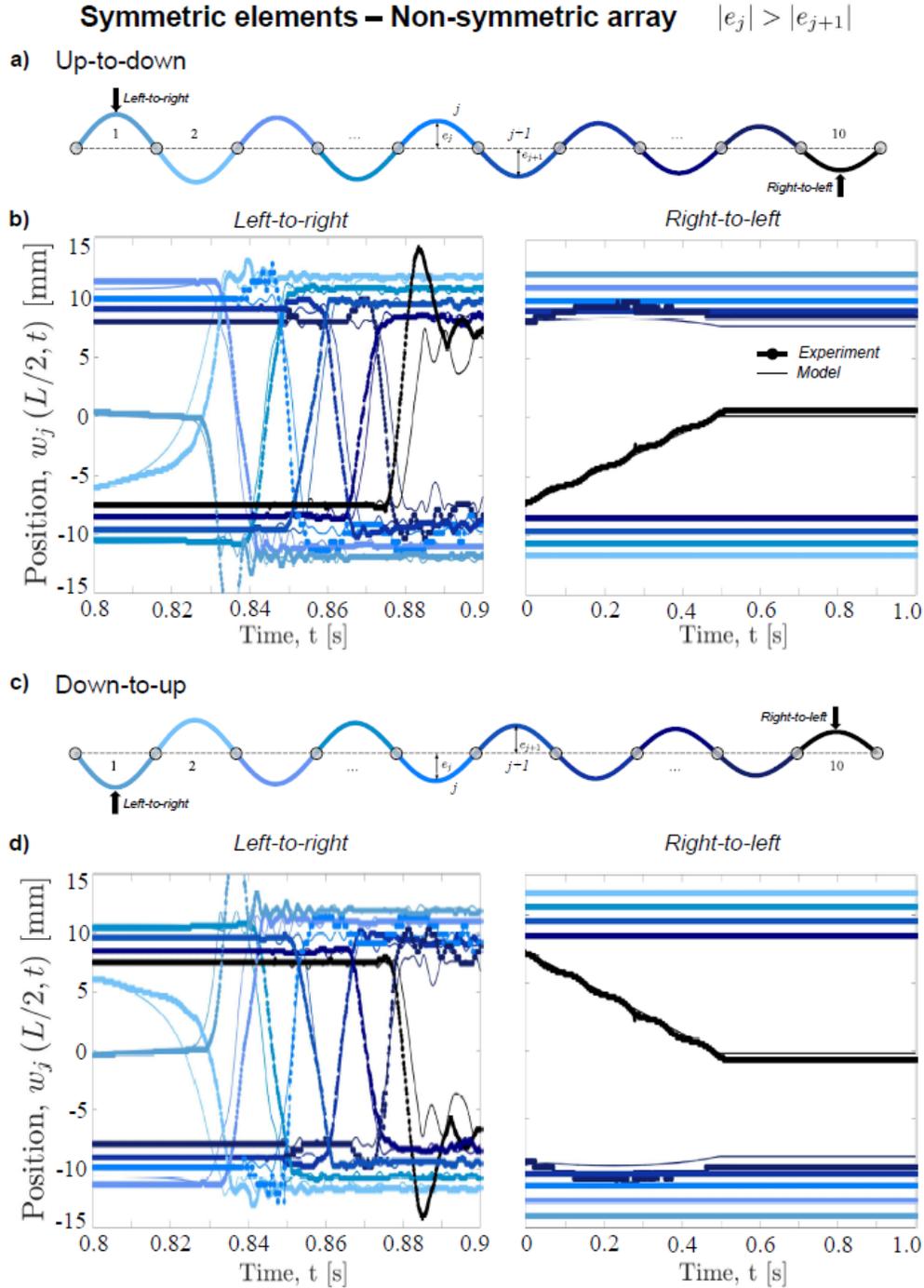

Figure S10: **Symmetric elements – Asymmetric array.** Results for the same structure considered in Fig. 3 of the main text. **a**, Schematic of the structure. **b**, Comparison between the experimentally measured (thick-dotted lines) and numerically predicted (thin lines) positions of the midpoints of the arches when the system is excited left-to-right and right-to-left as indicated in (a). **c**, Schematic of the structure. **d**, Comparison between the experimentally measured (thick-dotted lines) and numerically predicted (thin lines) positions of the midpoints of the arches when the system is excited left-to-right and right-to-left as indicated in (c).

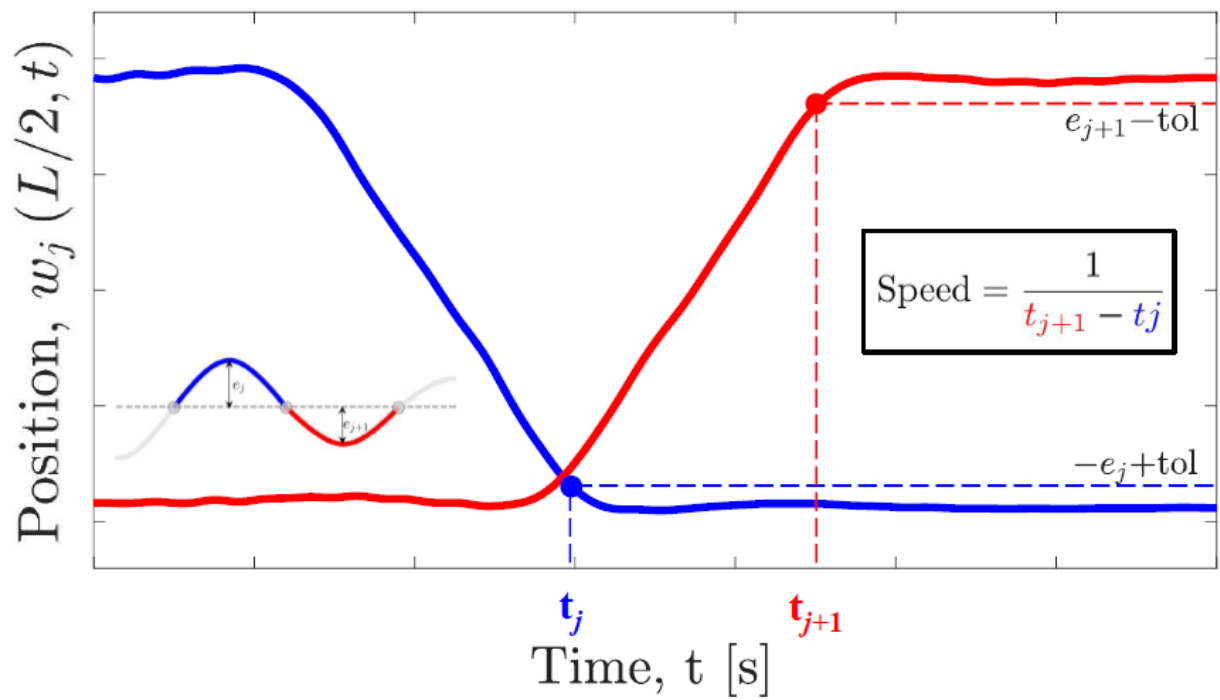

Figure S11: **Speed per unit.** Schematics illustrating how to compute the local speed in a 1D array of snapping arches.

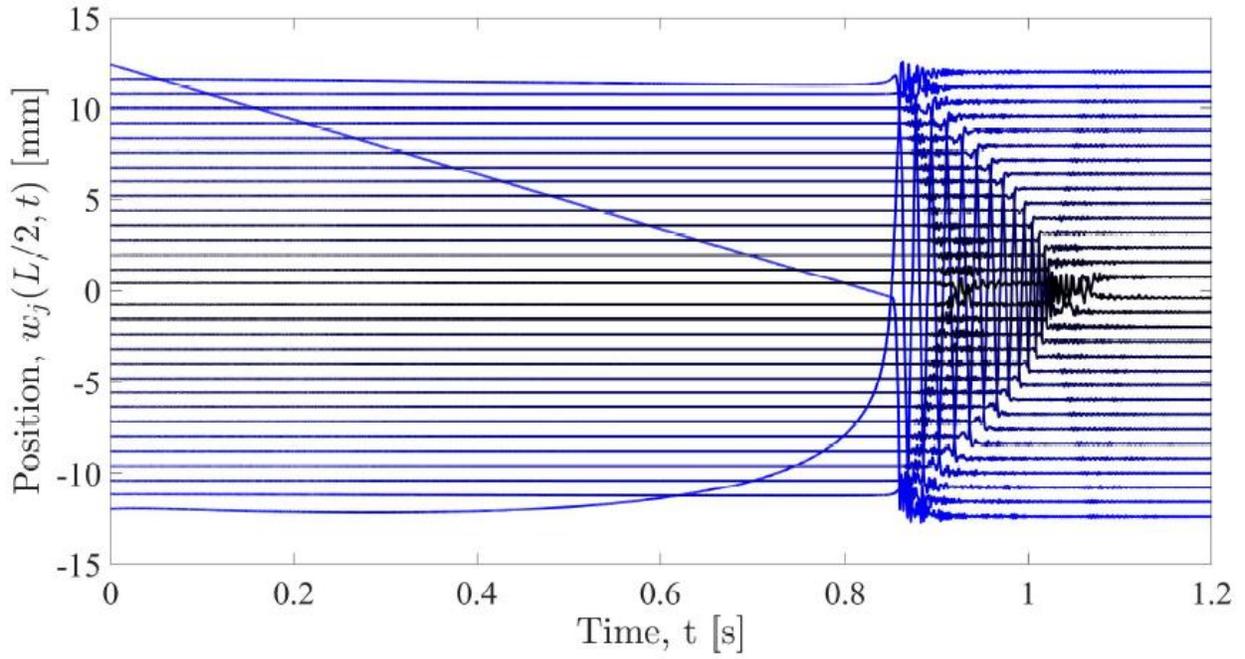

Figure S12: **Reversible diode.** Numerically predicted positions, $w_j(L/2, t)$, for an array comprising $N$ = 31 arches with modulated rise. For this chain $\Delta e_j$ = 400m and $e_1$ = 12.4 mm. The pulse is initiated by pushing down the leftmost arch.

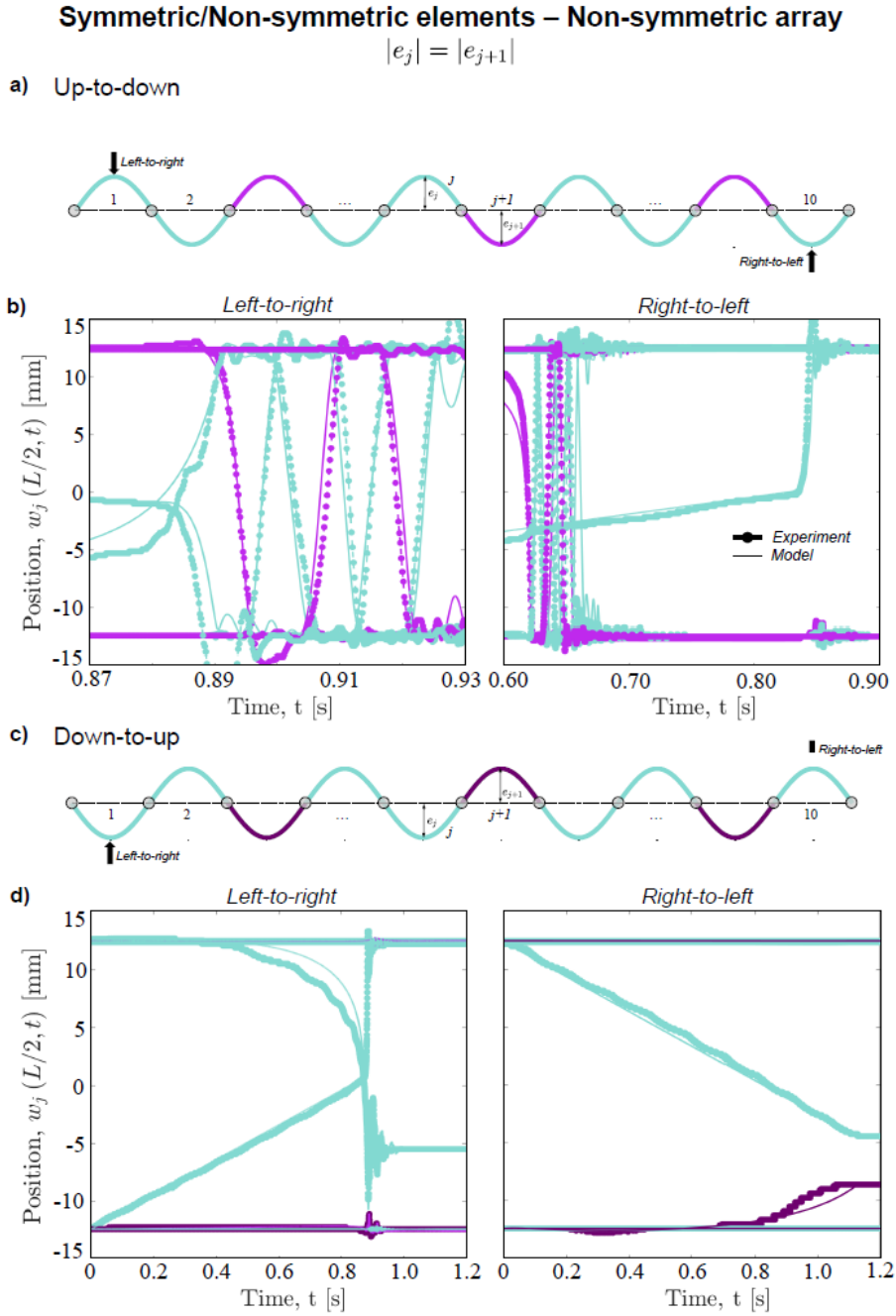

Figure S13: **Symmetric/Asymmetric elements – Asymmetric array.** Results for the same structure considered in Fig. 4 of the main text. **a**, Schematics of chain. **b**, Comparison between the experimentally measured (thick-dotted lines) and numerically predicted (thin lines) positions of the midpoints of the arches when the system is excited left-to-right and right-to-left as indicated in (a). **c**, Schematics of the chain. **d**, Comparison between the experimentally measured (thick-dotted lines) and numerically predicted (thin lines) positions of the midpoints of the arches when the system is excited left-to-right and right-to-left as indicated in (c). No transition waves are supported as indicated in (d).

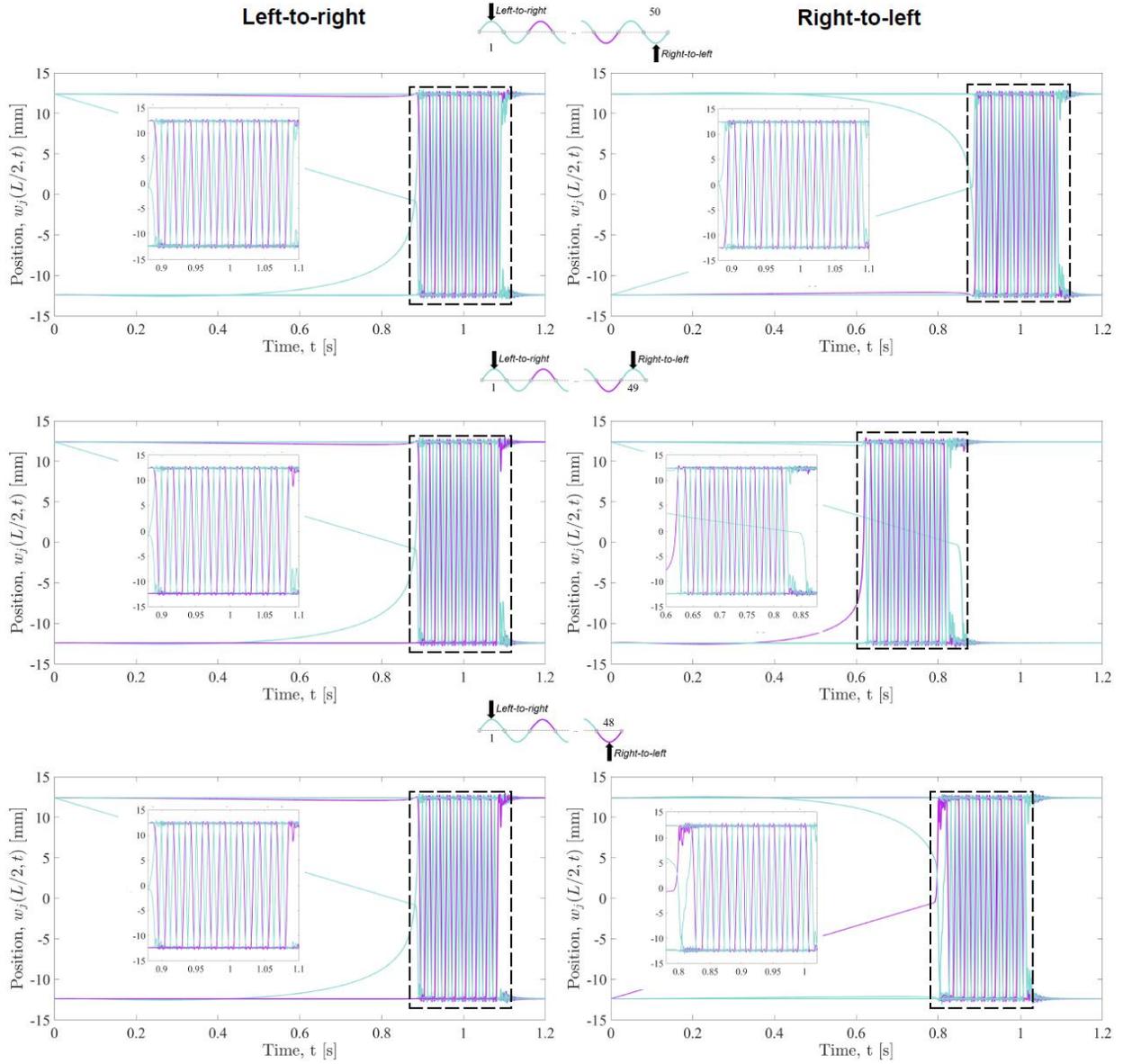

Figure S14: **Effect of chain symmetry/asymmetry – Tunable nonreciprocity** Numerically predicted positions of the arches' midpoints, $w_j(L/2, t)$, as a function of time for the three chains (comprising $N$ = 48, 49 and 50 arches) considered in Fig. 5a of the main text. Results for both left-to-right and right-to-left propagation are shown.

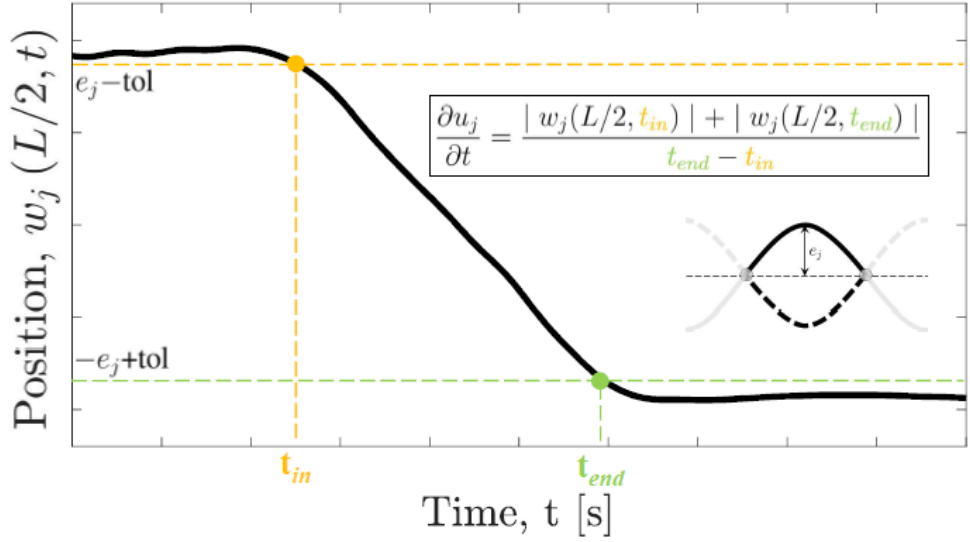

Figure S15: **Estimation of $c_{global}$ for a chain comprising only plastically deformed arches.** For a chain comprising only elements with asymmetric on-site energy potential the wave speed, $c_{global}$, can be estimated by balancing the total kinetic transported energy $E_d$, the difference $\Delta\varphi$ between the higher and lower energy well, and the energy dissipated as (*13*)

$$c_{global} = \frac{2\beta L E_d}{\Delta\phi} \tag{S23}$$

where

$$E_d = \sum_{j=1}^{N} \frac{1}{2}\left(\frac{\partial u_j}{\partial t}\right)^2 L \tag{S24}$$

with $\partial u_j/\partial t$ computed as showed in this figure. Using Eq. (S23), we find that in a chain comprising $N$ = 49 plastically deformed arches with $e_j$=12.4 mm $c_{global}$ ~ 503 units/s while from our simulations we compute a $c_{global}$ ~ 497 units/s, with a discrepancy around 1%.

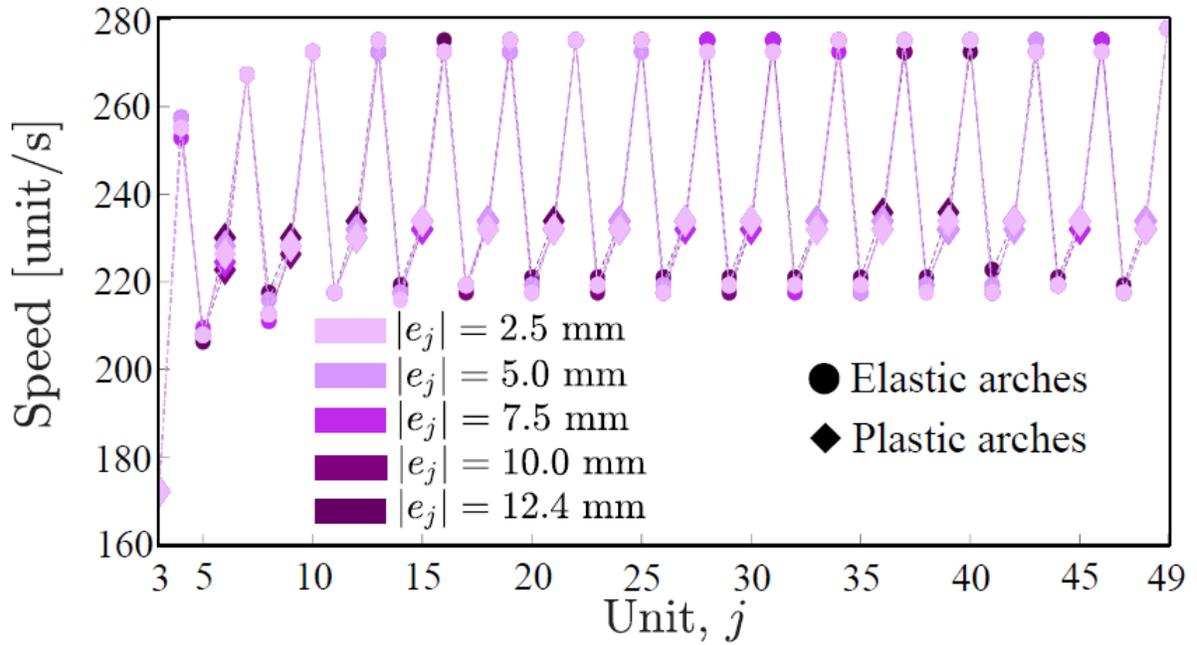

Figure S16: **Effect of the arch rise, $e_j$, on the response of a chain comprising both elastically and plastically deformed arches.** Numerically predicted local speed of the transition waves for different rises $e_j$ in a chain comprising $N = 49$ elastically and plastically deformed arches arranged as in Fig. 4b of the main text.

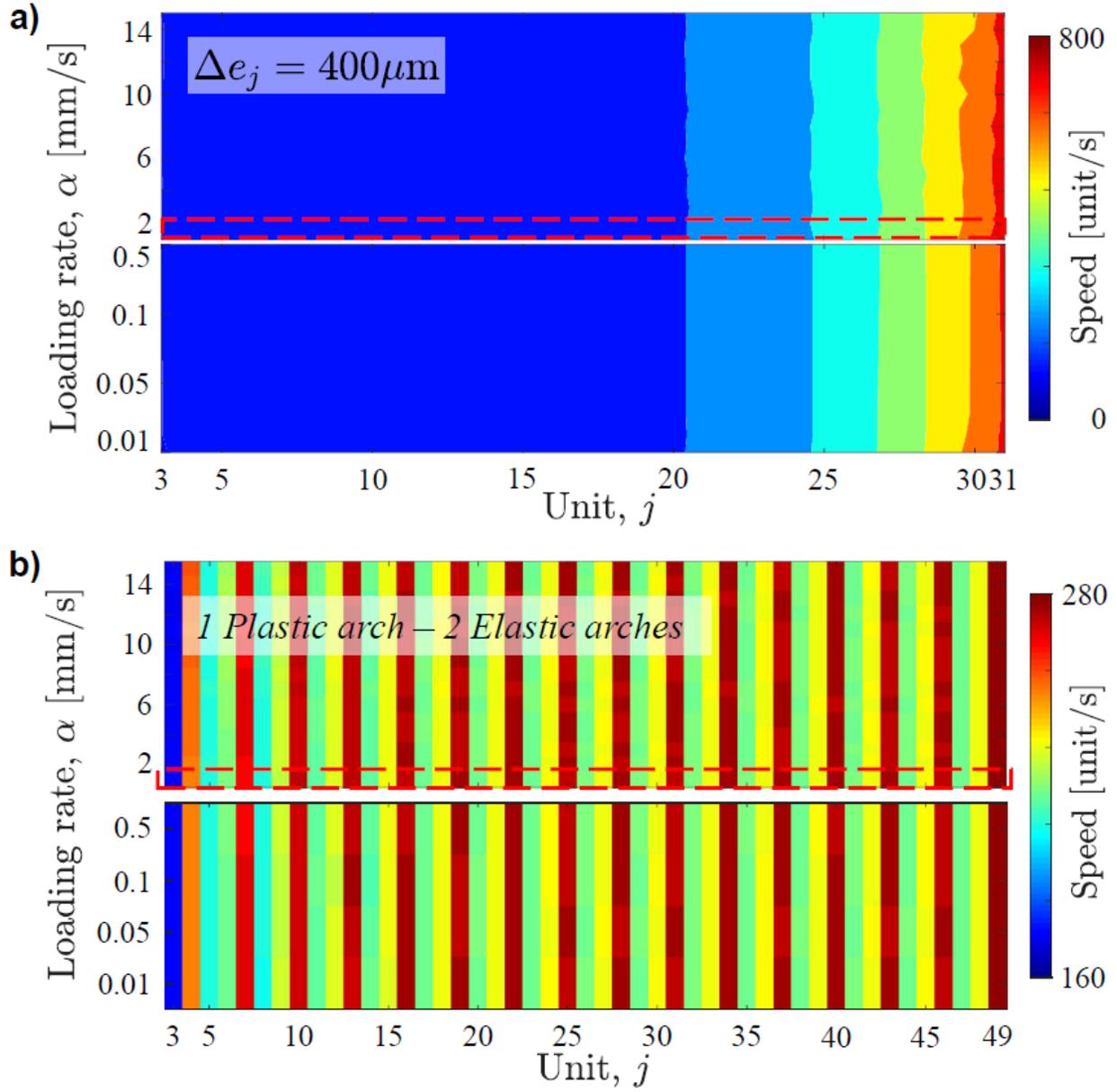

Figure S17: **Effect of loading rate** $\alpha$. **a**, Local speed of the transition waves in a graded system with $\Delta e_j = 400\mu$m and $e_1 = 12.4$ mm for different loading rates $\alpha$. We find that the wave speed is minimally affected by $\alpha$. **b**, Local speed of the transition waves for different loading rates $\alpha$ in a system comprising $N = 49$ elastically and plastically deformed arches with $e_j = 12.4$ mm arranged as in Fig. 4b of the main text. Also for this chain we find that the wave speed is minimally affected by $\alpha$.

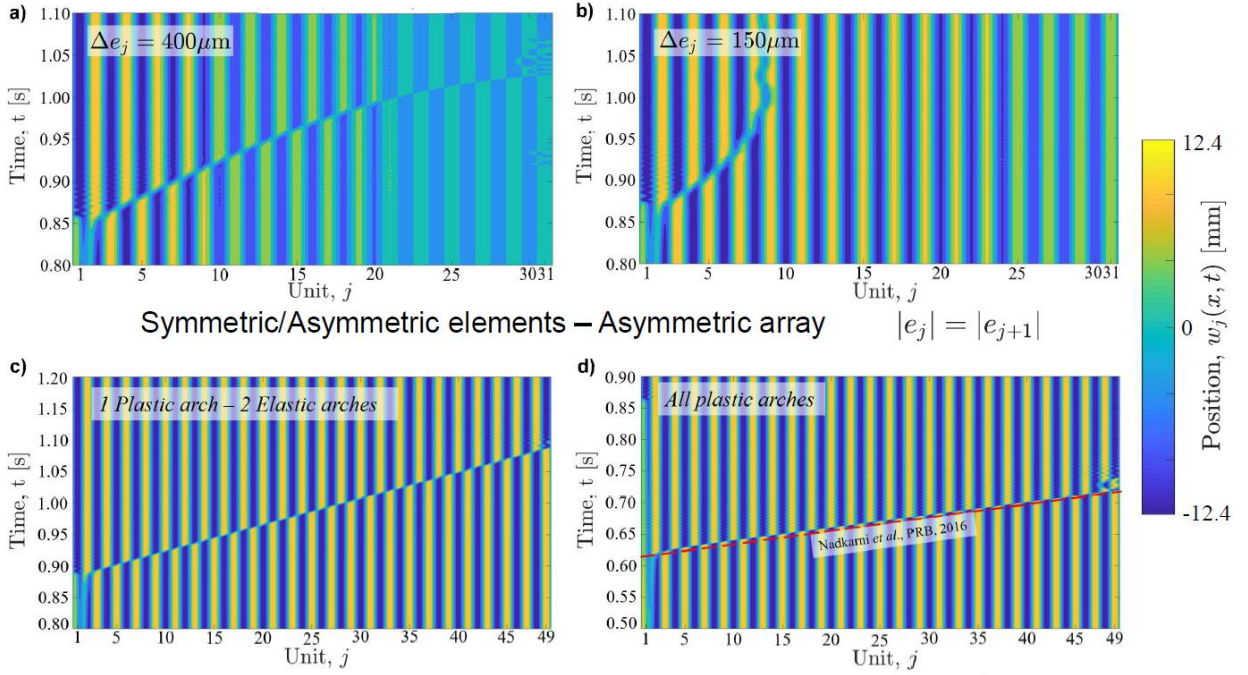

Figure S18: **Propagation of transition waves in different chains.** Numerically predicted position across each arch, $w_j(x_j, t)$, during the propagation of the wave in (a) a graded chain comprising elastically deformed arches with $\Delta e_j = 400$ $\mu$m and $e_1 = 12.4$ mm; (b) a graded chain comprising elastically deformed arches with $\Delta e_j = 150 \mu$m and $e_1 = 12.4$ mm; (c) a chain comprising $N = 49$ elastically and plastically deformed arches with $e_j = 12.4$ mm arranged as in Fig. 4b of the main text; (d) a chain comprising $N = 49$ plastically deformed arches with $e_j = 12.4$ mm. The red dashed line in (d) indicate the global speed, $c_{global}$, predicted by Eq. (S23).